\documentstyle[12pt]{article}
\begin{document}
\begin{titlepage}
\begin{center}

\hfill YITP-SB-02/26  \\
\vskip 20mm

{\Huge Four dimensional ${\cal R}^4$ superinvariants through gauge completion}

\vskip 10mm

Filipe Moura

\vskip 4mm

{\em C. N. Yang Institute for Theoretical Physics \\
State University of New York \\
Stony Brook, NY 11794-3840, U.S.A}

{\tt fmoura@insti.physics.sunysb.edu}

\vskip 6mm

\end{center}

\vskip .2in

\begin{center} {\bf Abstract } \end{center}
\begin{quotation}
\noindent
We fully compute the ${\cal N}=1$ supersymmetrization of the fourth power of 
the Weyl tensor in $d=4$ $x$-space with the auxiliary fields. In a previous 
paper, we showed that their elimination requires an infinite number of terms; 
we explicitely compute those terms to order $\kappa^4$ (three loop).
We also write, in superspace notation, all the possible ${\cal N}=1$
actions, in four dimensions, that contain pure ${\cal R}^4$ terms (with 
coupling constants). We explicitely write these actions in terms of the 
$\theta$ components of the chiral density $\epsilon$ and the supergravity 
superfields $R, G_m, W_{ABC}$. Using the method of gauge completion, we compute
the necessary $\theta$ components which allow us to write these actions in
$x$-space. We discuss under which circumstances can these extra ${\cal R}^4$ 
correction terms be reabsorbed in the pure supergravity action, and their 
relevance to the quantum supergravity/string theory effective actions.
\end{quotation}

\vfill
%%%%%%%%%%%%%%%%%%
\flushleft{\today}
%%%%%%%%%%%%%%%%%%

\end{titlepage}

\eject

%%%%%%%%%%%%%%%%%%%%%%%%%%%%%%%%%%%%%%%%%%%%%%%%%%%%%%%%%%%%%%%%%%%%%%
%%%%%%%%%%%%%%%%%%%%%%%%%%%%%%%%%%%%%%%%%%%%%%%%%%%%%%%%%%%%%%%%%%%%%%

\section{Introduction}
\indent
In a previous paper \cite{m01}, we wrote and analyzed in $d=4, {\cal N}=1$ 
superspace a lagrangian which contained a pure ${\cal R}^4$ term:
\begin{equation}
{\cal L}=\frac{1}{2 \kappa^2} \int E\left( 1+\alpha W^2\overline{W}^2\right) 
d^4\theta \label{action}
\end{equation}
This lagrangian could be seen as a four dimensional ${\cal N}=1$ 
string/M-theory effective action, resulting from a compactification and 
truncation from ten/eleven dimensions. In that case the constant 
$\frac{\alpha}{\kappa^2}$ could be identified up to a factor with 
$\alpha^{\prime 3}$, $\alpha^\prime$ being the string slope parameter.

Originally the $\alpha W^2\overline{W}^2$ term of this lagrangian was thought 
of as a possible three-loop quantum correction to supergravity. In reference 
\cite{dks77} it was shown that supersymmetry implies the one and two-loop
finiteness of supergravity, but there exists a potential non-vanishing 
on-shell superinvariant which could mean non-finiteness of supergravity at 
three loops. Later that superinvariant was written in superspace 
\cite{fz78,hl81}: 
it is precisely the $\alpha$ part of (\ref{action}). No one has been able to 
compute the three-loop supergravity effective action in order to show if the
coefficient of this term is nonzero or if it vanishes by some miraculous 
cancellation \cite{pvn81}. Up to that (unknown) numerical factor, the constant 
$\alpha$
could be identified with $\kappa^6$ if (\ref{action}) is that effective action.

In the paper \cite{m01}, we made a full superspace analysis of (\ref{action}).
In section \ref{2} of this paper, we extend this analysis to $x$-space. We 
write down (\ref{action}) in components, leaving the auxiliary fields. As 
we have shown in \cite{m01}, the auxiliary fields do not vanish anymore with 
this action; the elimination of the $A_m$ auxiliary field 
would result in a nonlocal action with an infinite number of terms of all 
orders in $\alpha$. We fully compute the terms which are of first order in 
$\alpha$ and which could constitute the three-loop supergravity effective 
action.

In section \ref{3} we argue that there are other supersymmetric
pure ${\cal R}^4$ terms which can be considered. These terms can be absorbed in
the pure supergravity action by a field redefinition, but this process induces 
new higher-order superinvariants.
We list these other supersymmetric ${\cal R}^4$ terms, compute their bosonic 
parts and comment on their properties.

%%%%%%%%%%%%%%%%%%%%%%%%%%%%%%%%%%%%%%%%%%%%%%%%%%%%%%%%%%%%%%%%%%%%%%
%%%%%%%%%%%%%%%%%%%%%%%%%%%%%%%%%%%%%%%%%%%%%%%%%%%%%%%%%%%%%%%%%%%%%%

\section{The ${\cal W}_+^2 {\cal W}_-^2$ supersymmetric action in $x$ space}
\label{2}
\indent
\setcounter{equation}{0} 
The lagrangian (\ref{action}) represents the supersymmetrization of one 
combination of ${\cal W}^4$, namely ${\cal W}_+^2 {\cal W}_-^2$ (for notation 
see appendix \ref{a1}). In order to write it in components, first we write it 
in chiral notation:
\begin{equation}
{\cal L}=\frac{1}{4 \kappa^2} \int \epsilon \left[\left( \overline{\nabla}^2 
+\frac{1}{3} \overline{R} \right) \left(-3 \alpha W^2\overline{W}^2 -3 \right) 
\right] d^2\theta + \mathrm{h.c.}
\end{equation}

$\epsilon$ is the chiral density; its expansion in components is 
\cite{lr79}
\begin{equation}
2\epsilon =e- ie \theta^A \sigma_{m A \dot A} \psi^{m \dot A} -e \theta^2 
\left(M-iN - \frac{1}{2} \psi^{m \dot A} \psi^{n \dot B} \sigma_{mn \dot A 
\dot B}\right)
\end{equation}

We get, then, for the lagrangian,

\begin{eqnarray}
\kappa^2 {\cal L}&=& -\frac{1}{2} e {\cal R} +\frac{1}{4} e 
\varepsilon^{\mu \nu \rho \lambda} \left( \psi_{\mu \dot A} 
\sigma_\nu^{A \dot A} \psi_{\rho \lambda A} - \psi_{\mu A} 
\sigma_\nu^{A \dot A} \psi_{\rho \lambda \dot A} \right) 
-\frac{1}{3} e \left(M^2 +N^2 - A^\mu A_\mu \right)
\nonumber \\
&-&\frac{3}{16} e \alpha \left. \nabla^2 W^2 \right|
\left. \overline{\nabla}^2 \overline{W}^2 \right|
-\frac{\alpha}{2} e {\cal R} \left. W^2 \right| 
\left. \overline{W}^2 \right| +3 \alpha e 
\partial_\mu \left. W^2 \right| \partial^\mu \left. \overline{W}^2 \right|
\nonumber \\
&+& \frac{\alpha}{4} e \left. W^2 \right| \left. \overline{W}^2 \right|
\varepsilon^{\mu \nu \rho \lambda} \left( \psi_{\mu \dot A} 
\sigma_\nu^{A \dot A} \psi_{\rho \lambda A} - \psi_{\mu A} 
\sigma_\nu^{A \dot A} \psi_{\rho \lambda \dot A} \right) \nonumber \\
&-& \frac{3}{4} i\alpha e  
\left( \left. \nabla^A W^2 \right| \sigma^\mu_{A \dot A} {\cal D}_\mu
\left. \nabla^{\dot A} \overline{W}^2 \right|+ 
\left. \nabla^{\dot A} \overline{W}^2 \right| \sigma^\mu_{A \dot A} 
{\cal D}_\mu \left. \nabla^A W^2 \right| \right) \nonumber \\
&+& \frac{3}{4} \alpha e \varepsilon^{\mu \nu \rho \lambda}
\psi_{\mu A} \sigma_\nu^{A \dot A} \psi_{\rho \dot A}
\left( \left. \overline{W}^2 \right| \partial_\lambda \left. W^2 \right| 
- \left. W^2 \right| \partial_\lambda \left. \overline{W}^2 \right| \right)
\nonumber \\
&-& \frac{3}{2} \alpha e \left( \psi^{\mu A} \left. \nabla_A W^2 \right| 
+ \sigma^{\mu \nu}_{AB} \psi_\nu^A \left. \nabla^B W^2 \right| \right)
\partial_\mu \left. \overline{W}^2 \right| \nonumber \\
&-& \frac{3}{2} \alpha e \left(\psi^{\mu \dot A} \left. \nabla_{\dot A} 
\overline{W}^2 \right|- \sigma^{\mu \nu}_{\dot A \dot B} \psi_\nu^{\dot A}
\left. \nabla^{\dot B} \overline{W}^2 \right| \right)
\partial_\mu \left. W^2 \right| \nonumber \\
&+& \frac{\alpha}{2} e \left(\left. \overline{W}^2 \right|
\sigma^{\mu \nu}_{AB} \left. \nabla^A W^2 \right| \psi_{\mu \nu}^B
+ \left. W^2 \right| \sigma^{\mu \nu}_{\dot A \dot B}
\left. \nabla^{\dot A} \overline{W}^2 \right| \psi_{\mu \nu}^{\dot B}
\right) \nonumber \\
&-& \frac{3}{4} \alpha e \left. \nabla^A W^2 \right| 
\left. \nabla^{\dot A} \overline{W}^2 \right| \left(
\psi_{\mu A} \psi^\mu_{\dot A} 
+\frac{1}{2} \sigma^{\mu \nu}_{\dot A \dot B} \psi_{\mu A} \psi_\nu^{\dot B}
+\frac{1}{2} \sigma^{\mu \nu}_{AB} \psi_\mu^B \psi_{\nu \dot A} \right)
\nonumber \\
&+& \frac{3}{8} \alpha e \varepsilon^{\mu \nu \rho \lambda} 
\sigma_{\lambda A \dot A} \left(\left. \overline{W}^2 \right| \psi_\mu^A
\psi_\nu^B \psi_\rho^{\dot A} \left. \nabla_B W^2 \right|
-\left. W^2 \right| \psi_\mu^A \psi_\nu^{\dot A} \psi_\rho^{\dot B}
\left. \nabla_{\dot B} \overline{W}^2 \right| \right) \nonumber \\
&+& \frac{3}{16} i\alpha e \left. W^2 \right| \psi^{\mu \dot B} 
\psi_{\mu \dot B} \psi_{\nu}^A \left. \nabla^{\dot A} \overline{W}^2 \right|
\sigma^\nu_{A \dot A} + \frac{3}{16} i\alpha e \left. \overline{W}^2 \right|
\psi^{\mu B} \psi_{\mu B} \psi_{\nu}^{\dot A} \left. \nabla^A W^2 \right|
\sigma^\nu_{A \dot A} \nonumber \\
&+& \frac{3}{8} i\alpha e \psi^{\mu A} \psi^{\dot A}_\mu \sigma^\nu_{A \dot A}
\left(\left. W^2 \right| \psi_{\nu}^{\dot B} \left. \nabla_{\dot B} 
\overline{W}^2 \right| - \left. \overline{W}^2 \right| \psi_{\nu}^B
\left. \nabla_B W^2 \right| \right) \nonumber \\
&-& \frac{3}{8} i\alpha e \sigma^\nu_{A \dot A} \left(\left. W^2 \right|
\psi^{\mu A} \psi^{\dot B}_\mu  \psi_{\nu \dot B} \left. \nabla^{\dot A} 
\overline{W}^2 \right| + \left. \overline{W}^2 \right| \psi_\nu^B \psi^\mu_B 
\psi^{\dot A}_\mu \left. \nabla^A W^2 \right| \right) \nonumber \\
&-& \frac{\alpha}{3} e \left(M^2 +N^2 - A^\mu A_\mu \right) 
\left. W^2 \right| \left. \overline{W}^2 \right| + \frac{\alpha}{4} e
\left. \nabla^A W^2 \right| \left. \nabla^{\dot A} \overline{W}^2 \right|
\sigma^\nu_{A \dot A} A_\nu \nonumber \\
&-& \frac{\alpha}{4} e 
\left(M+iN \right) \left. \overline{W}^2 \right| \left. \nabla^2 W^2 \right|
- \frac{\alpha}{4} e \left(M-iN \right) \left. W^2 \right|
\left. \overline{\nabla}^2 \overline{W}^2 \right| \nonumber \\
&+& i \alpha e \left( \left. \overline{W}^2 \right| \partial_\mu
\left. W^2 \right| - \left. W^2 \right| \partial_\mu \left. 
\overline{W}^2 \right| \right) A^\mu \nonumber \\
&-&\frac{i}{2} \alpha e \left( \left. \overline{W}^2 \right| 
\left. \nabla^A W^2 \right| \psi^\mu_A -\left. W^2 \right| \left. 
\nabla^{\dot A} \overline{W}^2 \right| \psi_{\dot A}^\mu \right) A_\mu
\label{xaction}
\end{eqnarray}

In order to compute this lagrangian in terms of the $x$-space fields, we are 
interested in the component expansion of $W^2$. From (\ref{w0}), one 
can derive
\begin{eqnarray}
\left. W^2 \right| &=& \frac{1}{6} \psi^{mnC} \psi_{mnC} -\frac{i}{24}
\varepsilon^{mnrs} \psi_{mn}^C \psi_{rsC} -\frac{i}{24} \varepsilon^{mnrt}
\psi_{mn}^A \psi_{rsC} \left( \sigma_t^{\ s} \right)_{\ \ A}^C \nonumber \\
&+& \frac{1}{12}A_n A^m \psi^{nA} \psi_{mA} -\frac{1}{12} A^m A_m \psi^{nA}
\psi_{nA} \nonumber \\
&-&\frac{i}{48} \varepsilon^{mnrt} A_s A_m \psi_n^A \psi_{rC} \left(
\sigma_t^{\ s} \right)^C_{\ \ A}  
+ \frac{i}{3} \psi^{mnA} A_m \psi_{nA} \nonumber \\
&+& \frac{1}{12} \varepsilon^{mnrs} A_r \psi_{mn}^C \psi_{sC}
+ \frac{1}{12} \varepsilon^{mnrt}
\psi_{mn}^A A_r \psi_{sC} \left( \sigma_t^{\ s} \right)_{\ \ A}^C
\end{eqnarray}

From the relation
\begin{equation}
\nabla_A W_{BCD}=\nabla_{\underline{A}} W_{\underline{BCD}} -\frac{1}{4}
\varepsilon_{AB} \nabla^E W_{ECD} 
-\frac{1}{4} \varepsilon_{AC} \nabla^E W_{EBD}
-\frac{1}{4} \varepsilon_{AD} \nabla^E W_{EBC} \label{da}
\end{equation}
one can easily show that one may 
write $\nabla_A W^2 =-2 W^{BCD} \nabla_A W_{BCD}$ as 
\begin{equation}
\nabla_A W^2 = \frac{1}{2} W_A^{\ \ BC} \nabla^D W_{DBC} - 2W^{BCD} 
\nabla_{\underline{A}} W_{\underline{BCD}} \label{dw2}
\end{equation}
and, using (\ref{diffw}), one can also show that
\begin{eqnarray}
\left. \nabla^A W_{ABC} \right| &=& -2i \sigma_{BC}^{mn} \left. 
\nabla_m G_n \right| \nonumber \\
&=& -\frac{2}{3} i \sigma_{BC}^{mn} e_m^{\ \ \mu} {\cal D}_\mu A_m 
+ i \sigma_{BC}^{mn} \psi_m^{\ \ A}
\left. \nabla_A G_n \right| 
+ i \sigma_{BC}^{mn} \psi_m^{\ \ \dot A} \left. \nabla_{\dot A} G_n \right|
\end{eqnarray}

In the same way one has, from (\ref{da}) and the solution to the Bianchi 
identities, 
\begin{eqnarray}
\nabla^2 W^2 &=& -2 \left(\nabla^A W^{BCD} \right) 
\nabla_A W_{BCD} +2 W^{ABC} \nabla^2 W_{ABC} \nonumber \\
&=& -2 \left(\nabla^A W^{BCD} \right) \nabla_{\underline{A}}
W_{\underline{BCD}} +12 \left(\nabla^m G^n \right) \nabla_m G_n \nonumber \\
&-& 12 \left(\nabla^m G^n \right) \nabla_n G_m -12i \varepsilon^{mnrs} 
\left(\nabla_m G_n \right) \nabla_r G_s  \nonumber \\
&-&\frac{5}{3} R W^2 - 20 W^{ABC} G_A^{\ \ \dot A} \nabla_B G_{C \dot A}
+8i W^{ABC} \nabla_A^{\ \ \dot A} \nabla_B G_{C \dot A} \label{d2w2}
\end{eqnarray}
with
\begin{eqnarray}
\left. \nabla_{\underline{A}}^{\ \ \dot A} \nabla_{\underline{B}} 
G_{\underline{C} \dot A} \right|&=& \sigma_{\underline{A}}^{m \ \dot A} 
e_m^{\ \mu} {\cal D}_\mu \left. \nabla_{\underline{B}} G_{\underline{C}\dot A}
\right| -2 \left(M-iN \right) \sigma_{\underline{A B}}^{mn}
\psi_{m \underline{C}} A_n \nonumber \\
&+& \frac{i}{6} \sigma_{\underline{A B}}^{mn} \psi_{m \underline{C}} 
e_n^{\ \mu} \partial_\mu \left(M-iN \right) - \frac{i}{48} 
\sigma_{\underline{A B}}^{mn} \psi_{m \underline{C}} \psi_n^{\dot D}
\left. \nabla_{\dot D} R \right| \nonumber \\
&+&\frac{i}{2} \sigma_{\underline{A B}}^{mn} \sigma_{\underline{C} \dot D}^p 
\psi_p^{\dot D} \left. \nabla_m G_n \right|
- \frac{1}{2} \sigma_{\underline{A}}^{m \ \dot A} \psi_m^{\dot D} 
\left. \nabla_{\underline{\dot D}} \nabla_{\underline{B}} 
G_{\underline{C} \underline{\dot A}} \right|
\end{eqnarray}

We have thus expressed the $x$-space  components necessary to compute our 
lagrangian in terms of the independent components listed in 
(\ref{independent}). These components, as well as 
$\left. \nabla_m G_n \right|$, are presented in appendix \ref{a1}. Some of them
are presented for the first time in the literature.
The knowledge of these components allows us to fully compute the lagrangian
(\ref{xaction}), leaving the auxiliary fields in it.

In \cite{m01}, we computed the field equations for (\ref{action}). We showed 
that the auxiliary fields $M, N$ satisfied an algebraic field equation, in 
terms of $e_\mu^{\ m}$, $\psi_\mu$ and the auxiliary field $A_m$. This field 
equation can be obtained by taking the $\theta=0$ component of 
\begin{equation}
R=6\alpha \overline{W}^2 \nabla^2 W^2+12\alpha^2
\overline{W}^4 W^2 \nabla^2 W^2
\end{equation}
These auxiliary fields can then be eliminated, leaving an action with higher 
powers of $\alpha$ and $A_m$.

The elimination of $A_m$ is much more problematic. Expressing this auxiliary 
field in terms of $e_\mu^{\ m}$ and $\psi_\mu$, we get a
series of terms with infinite powers of $\alpha$. This series cannot be 
expressed in a closed form; it is the solution of a differential equation - 
the field equation for $A_m$ - the solution of which can only be iterated order
by order in $\alpha$. Ingenuously, one would expect the effective action 
(\ref{action}) to be just of order $\kappa^4$ (or $\alpha^{\prime 3}$) but, 
from what we said, the closeness of the supersymmetry algebra 
requires, because of the elimination of $A_m$, the appearance of terms of 
infinitely high order in $\kappa$ (or $\alpha^\prime$). Notice that these 
terms all come from the auxiliary field sector and do not include terms which 
are pure (i.e. without matter couplings) powers of the Riemann tensor. These 
pure Riemann terms obviously exist in the full loop ($\alpha^\prime$) 
expansion, 
which is {\it not} fully given by (\ref{action}). The full quantum action 
will include more terms as (\ref{action}) but with larger powers of the
Riemann tensor. Each of them will {\it presumably} need, on-shell, an infinite
number of terms with all powers of the Riemann tensor to be supersymmetrized, 
like the simplest one we have analyzed.

The three-loop effective action (\ref{action}) was written based on the 
dimensional analysis of its leading bosonic part and on the requirement of 
supersymmetry. Since any regularization procedure works order by order in 
perturbation theory, it cannot introduce (at three loops or at any other loop) 
an infinite number of terms of arbitrary order in the coupling constant, as it 
is required by supersymmetry. Therefore, it is not possible to fully preserve 
supersymmetry while regulating the theory; supersymmetry transformations must 
be truncated to the power of $\kappa$ at which the theory is being regulated. 

Having this in mind, now we compute the three-loop (or $\alpha^{\prime 3}$) 
terms in $\left. \nabla^A W^2 \right|$ and $\left. \nabla^2 W^2 \right|$, i.e. 
the terms which would not contribute to (\ref{xaction}) 
with more powers of $\alpha$. Since all these terms are already multiplied by
$\alpha$, the terms we are looking for are simply the $\alpha=0$ (on-shell)
terms of $\left. \nabla_A W^2 \right|$ and $\left. \nabla^2 W^2 \right|$, 
i.e. the terms without auxiliary fields. From (\ref{dw2}), (\ref{d2w2}) and the
components in appendix \ref{a1}, it is only a matter of calculation. We leave
the details of the calculation to appendix \ref{a2}. Here, we present just the
results:
\begin{eqnarray}
\left. \nabla_A W^2 \right| &=& -\frac{1}{3} {\cal W}^+_{\mu \nu \rho \sigma}
\psi^{\mu \nu B} \sigma^{\rho \sigma}_{AB} 
+ \frac{i}{24} \sigma^{mn}_{A B} \sigma^{s D \dot A} 
\psi_{rs}^B \psi_{mn \dot A} \psi_D^r +\frac{i}{12}
\sigma^{ms}_{\underline{A} D} \sigma^r_{\underline{B} \dot A}
\psi_{\ s}^{n \ B} \psi_{mn}^{\dot A} \psi_r^D \nonumber \\
&-& \frac{i}{4} \sigma^r_{\underline{B} \dot A} \psi^{mn B} 
\psi_{mn}^{\dot A} \psi_{r \underline{A}}
-\frac{1}{8} \varepsilon^{mnrs} 
\sigma^{p}_{\underline{B} \dot A} \psi_{rs}^B \psi_{mn}^{\dot A} 
\psi_{p \underline{A}} +\frac{1}{12} \varepsilon_{mnrs}
\sigma^r_{\underline{B} \dot A} \psi^{p s B} 
\psi_p^{\ m \dot A} \psi^n_{\underline{A}} \nonumber \\
&-&\frac{1}{48} \varepsilon^{mnrs} \sigma^{pq}_{A B} \sigma_m^{D \dot A}  
\psi_{rs}^B \psi_{pq \dot A} \psi_{n D} + \frac{1}{24} \varepsilon^{mnrs}
\sigma^q_{\ s A D} \sigma^p_{\underline{B} \dot A}
\psi_{rq}^B \psi_{mn}^{\dot A} \psi_p^D \nonumber \\
&+&\frac{1}{24} \varepsilon^{mnrs} \sigma^{p}_{\underline{B} \dot A} 
\psi_{rp}^B \psi_{mn}^{\dot A} \psi_{s \underline{A}}
- \frac{1}{24} \varepsilon^{mnrs} \sigma_{s \underline{B} \dot A} 
\psi_{rp}^B \psi_{mn}^{\dot A} \psi^{p}_{\underline{A}} \nonumber \\
&+& \frac{i}{24} \sigma^{mn}_{A B} \sigma^{s D \dot A} 
\psi_{rs}^B \psi^r_{\dot A} \psi_{mn D} +\frac{i}{12}
\sigma^{ms}_{\underline{A} D} \sigma^r_{\underline{B} \dot A}
\psi_{\ s}^{n \ B} \psi_r^{\dot A} \psi^D_{mn}
- \frac{i}{4} \sigma^r_{\underline{B} \dot A} \psi_{mn}^B 
\psi_r^{\dot A} \psi^{mn}_{\underline{A}} \nonumber \\
&-&\frac{1}{8} \varepsilon^{mnrs} 
\sigma^{p}_{\underline{B} \dot A} \psi_{rs}^B \psi_p^{\dot A} 
\psi_{mn \underline{A}} +\frac{1}{12} \varepsilon_{mnrs}
\sigma^r_{\underline{B} \dot A} \psi^{p s B} 
\psi^{n \dot A} \psi^{\ m}_{p\ \underline{A}} \nonumber \\
&-&\frac{1}{48} \varepsilon^{mnrs} \sigma^{pq}_{A B} \sigma_m^{D \dot A} 
\psi_{rs}^B \psi_{n \dot A} \psi_{pq D} + \frac{1}{24} \varepsilon^{mnrs}
\sigma^q_{\ s A D} \sigma^p_{\underline{B} \dot A}
\psi_{rq}^B \psi_p^{\dot A} \psi_{mn}^D \nonumber \\
&+&\frac{1}{24} \varepsilon^{mnrs} \sigma^{p}_{\underline{B} \dot A} 
\psi_{rp}^B \psi_s^{\dot A} \psi_{mn \underline{A}}
- \frac{1}{24} \varepsilon^{mnrs} \sigma_{s \underline{B} \dot A} 
\psi_{rp}^B \psi^{p \dot A} \psi_{mn \underline{A}} 
+{\cal O}\left( \alpha \right) \label{dw20} \\
\left. \nabla^2 W^2 \right| &=& -2 {\cal W}^+_{\mu \nu \rho \sigma}
{\cal W}_+^{\mu \nu \rho \sigma} +\frac{8}{3}i {\cal W}^+_{\mu \nu \rho \sigma}
\psi^{\rho A} \psi^{\mu \nu \dot A} \sigma^\sigma_{A \dot A}
-\frac{8}{3}i {\cal W}^+_{\mu \nu \rho \sigma}
\psi^{\mu \nu A} \psi^{\rho \dot A} \sigma^\sigma_{A \dot A} \nonumber \\
&+&\frac{1}{8} \sigma^{mn}_{\underline{AB}} \sigma^{r\ \dot A}_{\underline{C}}
\sigma^{pq AB} \sigma^{s C \dot B} \psi_{r \underline{D}} \psi_{mn \dot A}
\psi_s^D \psi_{pq \dot B} \nonumber \\
&+&\frac{1}{8} \sigma^{mn}_{\underline{AB}} \sigma^{r\ \dot A}_{\underline{C}}
\sigma^{pq AB} \sigma^{s C \dot B} \psi_{mn \underline{D}} \psi_{r \dot A}
\psi_{pq}^D \psi_{s \dot B} \nonumber \\
&-&\frac{1}{4} \sigma^{mn}_{\underline{AB}} \sigma^{r\ \dot A}_{\underline{C}}
\sigma^{pq AB} \sigma^{s C \dot B} \psi_{r \underline{D}} \psi_{mn \dot A}
\psi_{pq}^D \psi_{s \dot B} +{\cal O}\left(\alpha \right) \label{d2w20}
\end{eqnarray}
The product $\sigma^{mn}_{AB} \sigma^{r\ \dot A}_C \sigma^{pq \underline{AB}}
\sigma^{s \underline{C} \dot B} \varepsilon^{\underline{D}E}$ is expanded in 
appendix \ref{a2}.

When replaced in (\ref{xaction}), these expressions will give all (and no more 
than that) terms which contribute with a power of $\kappa^4$ 
(or $\alpha^{\prime 3}$). This concludes our calculation of the three-loop
${\cal N}=1, d=4$ supergravity effective action, except for the numerical 
factor in the definition of $\alpha$.
%%%%%%%%%%%%%%%%%%%%%%%%%%%%%%%%%%%%%%%%%%%%%%%%%%%%%%%%%%%%%%%%%%%%%%
%%%%%%%%%%%%%%%%%%%%%%%%%%%%%%%%%%%%%%%%%%%%%%%%%%%%%%%%%%%%%%%%%%%%%%

\section{List and discussion of the different ${\cal R}^4$ superinvariants}
\label{3}
\setcounter{equation}{0}
\indent

As proven in \cite{fkwc92}, in four dimensions there are thirteen independent 
real scalar polynomials made from four powers of the irreducible components of 
the Riemann tensor, besides the ${\cal W}_+^2 {\cal W}_-^2$ which we analyzed 
in the previous section \footnote{We recall that the superfield 
components which contain the Riemann 
tensor are $\left. \nabla^2 \overline{R} \right|$ containing ${\cal R}$, 
$\left. \nabla_{\underline{A}} \nabla_{\underline{\dot A}} G_{\underline{B} 
\underline{\dot B}} \right|$ containing ${\cal S}_{\mu \nu} ={\cal R}_{\mu \nu}
 - \frac{1}{4} 
{\cal R} g_{\mu \nu}$, and $\left. \nabla_{\underline{A}} W_{\underline{BCD}} 
\right|$ containing ${\cal W}_{ABCD}:=-\frac{1}{8} {\cal W}^+_{\mu \nu \rho 
\sigma} \sigma^{\mu \nu}_{\underline{AB}} 
\sigma^{\rho \sigma}_{\underline{CD}}$ (see appendix \ref{a1}).}.
These terms are proportional to ${\cal R}$ or ${\cal S}_{\mu \nu}$ and 
therefore, when written in superspace language, they are proportional to $R$ or
$G_m$, with the single exception of ${\cal W}_+^4+{\cal W}_-^4$,
which we analyze at the end. 

For each fourth-order scalar polynomial of the Riemann tensor we now give its 
superspace supersymmetric form, when it exists. 
For all the component expansions see appendix \ref{a1}.
Also from these component expansions, or from the off-shell relations 
(\ref{diffg}), (\ref{diffw}), (\ref{wb203}), we can derive the $R$-weights of 
the superfields and their derivatives \cite{bgg01}:
\begin{equation}
\nabla_A \mapsto +1, R \mapsto +2, G_m \mapsto 0, W_{ABC} \mapsto -1
\end{equation}
Having these weights, we can mention in advance that all the actions we are 
going to consider preserve $R$-symmetry.

\subsection{${\cal R}^4$}
\indent

The supersymmetrization of the ${\cal R}^4$ polynomial is written, in 
superspace, as
\begin{equation}
\kappa^4 \int E R \overline{R} \left(\overline{\nabla}^2 R \right) 
\left(\nabla^2 \overline{R} \right) d^4\theta = -\frac{3}{2} \kappa^4
\int \epsilon \left( \overline{\nabla}^2 +\frac{1}{3} \overline{R} \right)
\left[R \overline{R} \left(\overline{\nabla}^2 R \right) 
\left(\nabla^2 \overline{R} \right) \right] d^2\theta + \mathrm{h.c.}
\end{equation}
This lagrangian contains a $\frac{3}{8} \kappa^4 e \left. \left(\nabla^2 
\overline{R} \right)^2 \right| \left. \left(\overline{\nabla}^2 R \right)^2
\right|$ term, the bosonic part of which being $1536 \kappa^4 e{\cal R}^4$.

\subsection{${\cal R}^2 {\cal S}_{\mu \nu} {\cal S}^{\mu \nu}$}
\indent

The supersymmetrization of the ${\cal R}^2 {\cal S}_{\mu \nu} 
{\cal S}^{\mu \nu}$ polynomial is included in the following superspace 
lagrangian:
\begin{equation}
\kappa^4 \int E G^2 \left(\overline{\nabla}^2 R \right) 
\left(\nabla^2 \overline{R} \right) d^4\theta= -\frac{3}{2} \kappa^4
\int \epsilon \left( \overline{\nabla}^2 +\frac{1}{3} \overline{R} \right)
\left[G^2 \left(\overline{\nabla}^2 R \right) 
\left(\nabla^2 \overline{R} \right) \right] d^2\theta + \mathrm{h.c.} 
\end{equation}
This lagrangian contains a $\frac{3}{16} \kappa^4 e \left. \nabla^2 
\overline{R}
\right| \left. \overline{\nabla}^2 R \right| \left. \left(\overline{\nabla}^2 
\nabla^2 + \nabla^2 \overline{\nabla}^2 \right) G^2 \right|$ term. From 
(\ref{d2g2}), and since $\left. \nabla^{\underline{A}} 
\nabla^{\underline{\dot A}} G^{\underline{B} \underline{\dot B}} \right|
\left. \nabla_A \nabla_{\dot A} G_{B \dot B} \right|= {\cal S}_{\mu \nu} 
{\cal S}^{\mu \nu}+\mbox{fermion terms,}$ 
$\left. \nabla^2 \overline{\nabla}^2 G^2 \right|
=-\frac{1}{18} {\cal R}^2 -2 {\cal S}_{\mu \nu} {\cal S}^{\mu \nu} 
+\mbox{fermion terms,}$ the bosonic part of this lagrangian is given by
$-\frac{4}{3} \kappa^4 e {\cal R}^4 -48 \kappa^4 e {\cal R}^2
{\cal S}_{\mu \nu} {\cal S}^{\mu \nu}$.
\subsection{${\cal R} {\cal S}_{\mu \nu} {\cal S}^{\mu}_{\ \sigma}
{\cal S}^{\nu \sigma}$}
\indent

The supersymmetrization of the ${\cal R} {\cal S}_{\mu \nu} 
{\cal S}^{\mu}_{\ \sigma} {\cal S}^{\nu \sigma}$ polynomial is written, in 
superspace, as
\begin{eqnarray}
&\kappa^4& \int E \left[R \left( \nabla_{\underline{A}} 
\nabla_{\underline{\dot A}} G_{\underline{B} \underline{\dot B}}\right) 
\left( \nabla^{\dot C} G^{A \dot A} \right) \left( \nabla_{\dot C} G^{B \dot B}
\right) + \mathrm{h.c.}\right] d^4\theta \nonumber \\
=& -&\frac{3}{2} \kappa^4
\int \epsilon \left( \overline{\nabla}^2 +\frac{1}{3} \overline{R} \right)
\left[R \left( \nabla_{\underline{A}} 
\nabla_{\underline{\dot A}} G_{\underline{B} \underline{\dot B}}\right) 
\left( \nabla^{\dot C} G^{A \dot A} \right) \left( \nabla_{\dot C} G^{B \dot B}
\right) + \mathrm{h.c.}\right] d^2\theta \nonumber \\
&+& \mathrm{h.c.} 
\end{eqnarray}
This action contains a $\frac{3}{16} \kappa^4 e \left. 
\overline{\nabla}^2 R \right| \left. \nabla_{\underline{A}} 
\nabla_{\underline{\dot A}} G_{\underline{B} \underline{\dot B}} \right|
\left. \nabla^C \nabla^{\dot C} G^{A \dot A} \right| \left. \nabla_C
\nabla_{\dot C} G^{B \dot B} \right| + \mathrm{h.c.}$ term. Since
$\left. \nabla_{\underline{A}} \nabla_{\underline{\dot A}} G_{\underline{B} 
\underline{\dot B}} \right| \left. \nabla^C \nabla^{\dot C} G^{A \dot A} 
\right| \left. \nabla_C \nabla_{\dot C} G^{B \dot B} \right|=
-{\cal S}_{\mu \nu} {\cal S}^{\mu}_{\ \sigma} {\cal S}^{\nu \sigma} 
+\mbox{fermion terms,}$ the bosonic part of this action is indeed
given by $3 \kappa^4 e {\cal R} {\cal S}_{\mu \nu} 
{\cal S}^{\mu}_{\ \sigma} {\cal S}^{\nu \sigma}$.

\subsection{${\cal S}_{\mu \nu} {\cal S}^{\mu \nu} {\cal S}_{\rho \sigma}
{\cal S}^{\rho \sigma}$}
\indent

The supersymmetrization of the ${\cal S}_{\mu \nu} {\cal S}^{\mu \nu} 
{\cal S}_{\rho \sigma} {\cal S}^{\rho \sigma}$ polynomial is included in the 
following superspace lagrangian:

\begin{equation}
\kappa^4 \int E \left(\overline{\nabla}^2 G^2 \right) 
\nabla^2 G^2 d^4\theta
\end{equation}
This lagrangian contains a $\frac{3}{8} \kappa^4 e \left. \nabla^2 
\overline{\nabla}^2 G^2 \right| \left. \overline{\nabla}^2 \nabla^2 G^2 
\right|$ term. Again from (\ref{d2g2}), the bosonic part of this lagrangian is
$\frac{3}{2} \kappa^4 e {\cal S}_{\mu \nu} {\cal S}^{\mu \nu} 
{\cal S}_{\rho \sigma} {\cal S}^{\rho \sigma} + \frac{1}{12}  \kappa^4 e
{\cal R}^2 {\cal S}_{\mu \nu} {\cal S}^{\mu \nu} + \frac{1}{432} \kappa^4 e 
{\cal R}^4$.

\subsection{${\cal S}_{\mu \nu} {\cal S}^{\mu}_{\ \sigma}
{\cal S}^{\rho \sigma} {\cal S}^{\nu}_{\ \rho}$}
\indent

The supersymmetrization of the ${\cal S}_{\mu \nu} {\cal S}^{\mu}_{\ \sigma}
{\cal S}^{\rho \sigma} {\cal S}^{\nu}_{\ \rho}$ polynomial is written, in superspace, as
\begin{eqnarray}
&\kappa^4& \int E \left( \nabla^{\dot C} G^{A \dot A} \right) 
\left( \nabla_{\dot C} G^{B \dot B} \right) \left( \nabla^C G_{A \dot A} 
\right) \left( \nabla_C G^{B \dot B} \right) d^4\theta \nonumber \\
= &-&\frac{3}{2} \kappa^4
\int \epsilon \left( \overline{\nabla}^2 +\frac{1}{3} \overline{R} \right)
\left[\left( \nabla^{\dot C} G^{A \dot A} \right) 
\left( \nabla_{\dot C} G^{B \dot B} \right) \left( \nabla^C G_{A \dot A} 
\right) \left( \nabla_C G^{B \dot B} \right) \right] d^2\theta \nonumber \\
&+& \mathrm{h.c.} 
\end{eqnarray}
This action contains a $\frac{3}{8} \kappa^4 e \left. 
\nabla^{\underline{A}} 
\nabla^{\underline{\dot A}} G^{\underline{C} \underline{\dot C}} \right|
\left. \nabla_A \nabla_{\dot A} G^{D \dot D} \right| \left. \nabla^{\dot B}
\nabla^B G_{C \dot C} \right| \left. \nabla_{\underline{\dot B}} 
\nabla_{\underline{B}} G_{\underline{D} \underline{\dot D}} \right|$ term,
the bosonic part of which being given by 
$\frac{3}{8} \kappa^4 e {\cal S}_{\mu \nu} {\cal S}^{\mu}_{\ \sigma}
{\cal S}^{\rho \sigma} {\cal S}^{\nu}_{\ \rho}$.

\subsection{${\cal R} {\cal S}_{\mu \rho} {\cal S}_{\nu \sigma}
{\cal W}^{\mu \nu \rho \sigma}$}
\indent

The supersymmetrization of the ${\cal R} {\cal S}_{\mu \rho} 
{\cal S}_{\nu \sigma} {\cal W}^{\mu \nu \rho \sigma}$ polynomial is written, 
in superspace, as
\begin{eqnarray}
&\kappa^4& \int E \left [\overline{R} \left( \nabla^A G^{B \dot B} \right)
\left( \nabla^C G^D_{\ \dot B} \right) \nabla _D W_{ABC} - R
\left( \nabla^{\dot A} G^{B \dot B} \right) \left( \nabla^{\dot C} 
G_B^{\ \dot D} \right) \nabla_{\dot D} W_{\dot A \dot B \dot C}
\right] d^4\theta \nonumber \\
= &-&\frac{3}{2} \kappa^4
\int \epsilon \left( \overline{\nabla}^2 +\frac{1}{3} \overline{R} \right)
\left[\overline{R} \left( \nabla^A G^{B \dot B} \right)
\left( \nabla^C G^D_{\ \dot B} \right) \nabla_D W_{ABC} + \mathrm{h.c.}\right] 
d^2\theta + \mathrm{h.c.}
\end{eqnarray}
This lagrangian contains a $\frac{3}{16}\kappa^4 e \left. \nabla^2 \overline{R}
\right| \left. \nabla^{\dot A} \nabla^A G^{B \dot B} \right|
\left. \nabla_{\underline{\dot A}} \nabla^C G^D_{\ \underline{\dot B}} \right|
\left. \nabla_{\underline{A}} W_{\underline{BCD}} \right| + \mathrm{h.c.} $
term. Using ({\ref{sss}) one can derive 
$\left. \nabla^{\dot A} \nabla^A G^{B \dot B} \right|
\left. \nabla_{\underline{\dot A}} \nabla^C G^D_{\ \underline{\dot B}} \right|
\left. \nabla_{\underline{A}} W_{\underline{BCD}} \right| = -\frac{4}{3}
{\cal S}_{\mu \rho} {\cal S}_{\nu \sigma} {\cal W}_+^{\mu \nu \rho \sigma}
+\mbox{fermionic terms}$, and one can then show that
the pure bosonic part of this lagrangian is $2 \kappa^4 e {\cal R} 
{\cal S}_{\mu \rho} {\cal S}_{\nu \sigma} {\cal W}^{\mu \nu \rho \sigma}$.

\subsection{${\cal R}^2 {\cal W}_{\mu \nu \rho \sigma}
{\cal W}^{\mu \nu \rho \sigma}$}
\indent

The supersymmetrization of the ${\cal R}^2 {\cal W}_{\mu \nu \rho \sigma}
{\cal W}^{\mu \nu \rho \sigma}$ polynomial is written, in superspace, as
\begin{eqnarray}
&\kappa^4& \int E R \overline{R} \left(\nabla^2 W^2 + \overline{\nabla}^2 
\overline{W}^2 \right) d^4\theta \nonumber \\
= &-&\frac{3}{2} \kappa^4
\int \epsilon \left( \overline{\nabla}^2 +\frac{1}{3} \overline{R} \right)
\left[R \overline{R} \left(\overline{\nabla}^2 \overline{W}^2 
+ \nabla^2 W^2 \right) \right] d^2\theta + \mathrm{h.c.} \label{s2w20}
\end{eqnarray}
This lagrangian contains a $\frac{3}{8} \kappa^4 e \left. \nabla^2 \overline{R}
\right| \left. \overline{\nabla}^2 R \right| \left. \left( \nabla^2 W^2 
+\overline{\nabla}^2 \overline{W}^2 \right) \right|$ term, the bosonic part of 
which being given by $-48 \kappa^4 e {\cal R}^2 {\cal W}_{\mu \nu \rho \sigma}
{\cal W}^{\mu \nu \rho \sigma}$.

\subsection{${\cal S}_{\tau \lambda} {\cal S}^{\tau \lambda} 
{\cal W}_{\mu \nu \rho \sigma} {\cal W}^{\mu \nu \rho \sigma}$}
\indent

The supersymmetrization of the ${\cal S}_{\tau \lambda} {\cal S}^{\tau \lambda}
{\cal W}_{\mu \nu \rho \sigma} {\cal W}^{\mu \nu \rho \sigma}$ polynomial
is included in the following superspace lagrangian:

\begin{eqnarray}
&\kappa^4& \int E G^2 \left(\nabla^2 W^2 + \overline{\nabla}^2 \overline{W}^2 
\right) d^4\theta \nonumber \\
= &-&\frac{3}{2} \kappa^4
\int \epsilon \left( \overline{\nabla}^2 +\frac{1}{3} \overline{R} \right)
\left[G^2 \left(\overline{\nabla}^2 \overline{W}^2 
+ \nabla^2 W^2 \right) \right] d^2\theta + \mathrm{h.c.} \label{s2w21}
\end{eqnarray}
This lagrangian contains a $\frac{3}{16} \kappa^4 e \left. 
\left(\overline{\nabla}^2 \nabla^2 + \nabla^2 \overline{\nabla}^2 \right) G^2 
\right| \left. \left( \nabla^2 W^2 +\overline{\nabla}^2 \overline{W}^2 \right) 
\right|$ term. Its pure bosonic part is given by $\frac{1}{24} 
\kappa^4 e {\cal R}^2 {\cal W}_{\mu \nu \rho \sigma} 
{\cal W}^{\mu \nu \rho \sigma}+\frac{3}{2} e \kappa^4 
{\cal S}_{\tau \lambda} {\cal S}^{\tau \lambda}
{\cal W}_{\mu \nu \rho \sigma} {\cal W}^{\mu \nu \rho \sigma}$.

\subsection{${\cal S}_{\mu \nu} {\cal S}^{\tau \lambda} 
{\cal W}_{\tau \rho \lambda \sigma} {\cal W}^{\mu \rho \nu \sigma}$}
\indent

The supersymmetrization of the ${\cal S}_{\mu \nu} {\cal S}^{\tau \lambda} 
{\cal W}_{\tau \rho \lambda \sigma} {\cal W}^{\mu \rho \nu \sigma}$ 
polynomial contains the following superspace lagrangian:

\begin{eqnarray}
&\kappa^4& \int E W_{BCD} W_{\dot B \dot C \dot D} \left(\nabla^{\dot B} 
G^{B \dot C} \right)
\nabla^C G^{D \dot D} d^4\theta \nonumber \\
&=& -\frac{3}{2} \kappa^4 \int \epsilon \left( \overline{\nabla}^2 +\frac{1}{3}
\overline{R} \right) \left(W_{BCD} W_{\dot B \dot C \dot D} 
\left(\nabla^{\dot B} G^{B \dot C} \right) \nabla^C G^{D \dot D} \right)
d^2\theta +\mathrm{h.c.}
\end{eqnarray}
This lagrangian contains a $\frac{3}{8} \kappa^4 e \left. 
\nabla_{\underline{A}} W_{\underline{BCD}} \right|
\left. \nabla_{\underline{\dot A}} W_{\underline{\dot B \dot C \dot D}} \right|
\left. \nabla^A \nabla^{\dot B} G^{B \dot C} \right|
\left. \nabla^{\dot A} \nabla^C G^{D \dot D} \right|$ term. Since
$\left. \nabla_{\underline{A}} W_{\underline{BCD}} \right| \left. \nabla^A 
\nabla_{\underline{\dot A}} G^B_{\ \underline{\dot B}} \right|=\frac{1}{12} 
{\cal S}^{\mu}_{\ \tau}
{\cal W}^+_{\mu \nu \rho \sigma} \sigma^{\rho \sigma}_{CD}
\sigma^{\tau \nu}_{\dot A \dot B} +\frac{2}{3} {\cal W}^+_{\mu \nu \rho \sigma}
{\cal S}^{\mu \rho} \sigma^\nu_{\underline{C} \underline{\dot A}} 
\sigma^\sigma_{\underline{D} \underline{\dot B}}+\mbox{fermionic}$ terms, 
using the symmetries of the Weyl tensor one can derive the pure bosonic part of
this lagrangian, which is $-\frac{4}{3} \kappa^4 e 
{\cal S}_{\mu \nu} {\cal S}^{\tau \lambda} 
{\cal W}^+_{\tau \rho \lambda \sigma} {\cal W}_-^{\mu \rho \nu \sigma}$.

\subsection{${\cal S}_{\mu}^{\ \tau} {\cal S}_{\nu}^{\ \lambda} 
{\cal W}_{\tau \lambda \rho \sigma} {\cal W}^{\mu \nu \rho \sigma}$}
\indent

The supersymmetrization of the ${\cal S}_{\mu}^{\ \tau} 
{\cal S}_{\nu}^{\ \lambda} 
{\cal W}_{\tau \lambda \rho \sigma} {\cal W}^{\mu \nu \rho \sigma}$ polynomial
is written, in superspace, as
\begin{eqnarray}
&\kappa^4& \int E \left[W_{ABC} W^{CEF} \left( \nabla^A G_{E \dot B} \right)
\nabla^B G_F^{\ \dot B} + \mathrm{h.c.} \right] d^4\theta \nonumber \\
= &-&\frac{3}{2} \kappa^4 \int \epsilon \left( \overline{\nabla}^2 +\frac{1}{3}
\overline{R} \right) \left[W_{ABC} W^{CEF} \left( \nabla^A G_{E \dot B} \right)
\nabla^B G_F^{\ \dot B} + \mathrm{h.c.} \right] d^2\theta \nonumber \\
&+& \mathrm{h.c.} \label{s2w22}
\end{eqnarray}
This lagrangian contains a $-\frac{3}{16} \kappa^4 e \left. 
\nabla_{\underline{A}} W_{\underline{BCD}} \right| \left. \nabla^C W^{DEF} 
\right| \left. \nabla^{\underline{\dot A}} \nabla^A G_{E \underline{\dot B}}
\right| \left. \nabla_{\dot A} \nabla^B G_F^{\ \dot B} \right|+ \mathrm{h.c.}$ 
term. Since ${\cal W}^{CDBF} {\cal W}_{CDAE}= \frac{1}{8} \sigma^{\mu \nu}_{BF}
\sigma^{AE}_{\tau \lambda} {\cal W}^+_{\mu \nu \rho \sigma}
{\cal W}_+^{\tau \lambda \rho \sigma}$, the pure bosonic part of this 
lagrangian is easily shown to be $\frac{3}{8} \kappa^4 e 
{\cal W}_{\tau \lambda \rho \sigma} {\cal W}^{\mu \nu \rho \sigma}
{\cal S}_{\mu}^{\ \tau} {\cal S}_{\nu}^{\ \lambda}$.

\subsection{${\cal R} {\cal W}_{\mu \nu}^{\ \ \tau \lambda} 
{\cal W}_{\tau \lambda \rho \sigma} {\cal W}^{\mu \nu \rho \sigma}$}
\indent

The supersymmetrization of the ${\cal R} {\cal W}_{\mu \nu}^{\ \ \tau \lambda} 
{\cal W}_{\tau \lambda \rho \sigma} {\cal W}^{\mu \nu \rho \sigma}$ polynomial
is written, in superspace, as
\begin{eqnarray}
&\kappa^4& \int E \left(R  W^{ABC}  W_A^{\ DE} \nabla_B W_{CDE} + \mathrm{h.c.}
\right) d^4\theta \nonumber \\ 
= &-&\frac{3}{2} \kappa^4 \int \epsilon \left( \overline{\nabla}^2 +\frac{1}{3}\overline{R} \right) \left(R  W^{ABC}  W_A^{\ DE} \nabla_B W_{CDE} + 
\mathrm{h.c.} \right) d^2\theta +\mathrm{h.c.} \label{rwww}
\end{eqnarray}
This lagrangian contains a $\frac{3}{16} \kappa^4 e \left. \overline{\nabla}^2 
R \right| \left. \nabla^F W^{ABC} \right| \left. \nabla_F W_A^{\ DE} \right| 
\left. \nabla_B W_{CDE} \right| +\mathrm{h.c.}$ term. Since ${\cal W}^{ABCF} 
{\cal W}_{AF}^{\ \ \ DE} {\cal W}_{BCDE} =-\frac{4}{9} 
{\cal W}_{\mu \nu}^{+ \ \tau \lambda} 
{\cal W}^+_{\tau \lambda \rho \sigma} {\cal W}_+^{\mu \nu \rho \sigma}$, the 
pure bosonic part of this lagrangian is $\frac{2}{3} \kappa^4 e {\cal R} 
{\cal W}_{\mu \nu}^{\ \ \tau \lambda} 
{\cal W}_{\tau \lambda \rho \sigma} {\cal W}^{\mu \nu \rho \sigma}$.

\subsection{Discussion of the results}
\indent

We have identified, for a total of twelve independent real scalar polynomials 
made from four powers of irreducible components of the Riemann tensor, their
corresponding superspace lagrangians. Obviously our choice of basis for these 
polynomials is not unique, and linear combinations of the superspace 
lagrangians we found can be taken freely, in order to supersymmetrize any 
desired linear combination of the Riemann polynomials.

Also other contractions of indices could have been taken, both in the Riemann 
polynomials and in the superspace lagrangians, but they would always be 
equivalent to some linear combination of the independent lagrangians we chose. 
For example, let's consider 

\begin{eqnarray}
&\kappa^4& \int E \left[W_{BCD} W^{BEF} \left( \nabla_E G_{F \dot B} \right)
\nabla^C G^{D \dot B} + \mathrm{h.c.} \right] d^4\theta \nonumber \\
= &-&\frac{3}{2} \kappa^4 \int \epsilon \left( \overline{\nabla}^2 +\frac{1}{3}
\overline{R} \right) \left[W_{BCD} W^{BEF} \left( \nabla_E G_{F \dot B} \right)
\nabla^C G^{D \dot B} + \mathrm{h.c.} \right] d^2\theta \nonumber \\
&+& \mathrm{h.c.} \label{s2w23}
\end{eqnarray}
This lagrangian contains a $\frac{3}{16} \kappa^4 e \left. 
\nabla_{\underline{A}} W_{\underline{BCD}} \right| \left. \nabla^A W^{BEF} 
\right| \left. \nabla_{\underline{\dot A}} \nabla_E G_{F \underline{\dot B}}
\right| \left. \nabla^{\dot A} \nabla^C G^{D \dot B} \right|+ \mathrm{h.c.}$ 
term. From our previous results, it is easy to show that its pure bosonic part
is given by
$\frac{1}{24} \kappa^4 e {\cal S}_{\mu}^{\ \tau} {\cal S}_{\nu}^{\ \lambda} 
{\cal W}_{\tau \lambda \rho \sigma} {\cal W}^{\mu \nu \rho \sigma} + 
\frac{1}{96} \kappa^4 e {\cal S}_{\tau \lambda} {\cal S}^{\tau \lambda} 
{\cal W}_{\mu \nu \rho \sigma} {\cal W}^{\mu \nu \rho \sigma}$. It is easy to
see that it is a linear combination of the bosonic parts of (\ref{s2w20}), 
(\ref{s2w21}) and (\ref{s2w22}). Indeed, by partial integration in superspace, 
one can see directly that (\ref{s2w23}) is a linear combination of 
(\ref{s2w20}), (\ref{s2w21}) and (\ref{s2w22}).

All the supersymmetrizations we have been considering, as we mentioned, are 
proportional to $R$ or $G_m$. If they are written simply as quantum corrections
to pure 
supergravity, they can be reabsorbed by field redefinitions, according to 
\cite{dks77, hv74}. In superspace this can be seen very clearly. We recall that
the general variation of the supergravity action 
$I=\frac{1}{2 \kappa^2} \int \int E d^4\theta d^4x $ under any transformation
of the supervielbein which preserves the off-shell torsion constraints is given
by \cite{wz781}

\begin{equation}
2 \kappa^2
\delta I=\int E \left( \frac{2}{3} i\chi^{A \dot A} G_{A \dot A}
+\frac{1}{9} \overline{R} \overline{U} + \frac{1}{9} R U \right) d^4x
d^4\theta \label{deltapure2}
\end{equation}
The superfields $U, \chi^{A \dot A}$ are completely arbitrary. Any correction 
term proportional to $R$ or $G_m$ can be written in the form (\ref{deltapure2})
and, therefore, reabsorbed in the supergravity action by a redefinition of
the supervielbein. In fact, these correction terms to pure supergravity vanish 
on shell, because the field equations for pure supergravity are $R=0$ and 
$G_m=0$. The supervielbein redefinition will, though, introduce new higher 
order superinvariants (by higher order, in this section, we mean $\kappa^{10}$
or more), which may vanish on-shell or not, as we will see below.

The correction in (\ref{action}) cannot be written in the form 
(\ref{deltapure2}), nor can its variation. In this case, the variation is 
much more complicated and, although still being proportional to the arbitrary 
superfields 
$U, \chi^{A \dot A}$, it includes terms with derivatives of $R$, $G_m$ and
$W_{ABC}$ \cite{m01}. These terms imply the nontrivial field equations for the
auxiliary fields.

There is an interesting aspect about the 
${\cal R} {\cal W}_{\mu \nu}^{\ \ \tau \lambda} 
{\cal W}_{\tau \lambda \rho \sigma} {\cal W}^{\mu \nu \rho \sigma}$
action (\ref{rwww}): it has three powers of the
Weyl tensor, while all the other actions we have been considering in this 
section have at
most two powers of the Weyl tensor. This means all the previous actions have, 
in superspace, up to derivatives, at least either two $G_m$ factors, or one 
$R \overline{R}$ factor, or one $(R +\overline{R}) G_m$ factor. It is then 
obvious that these actions will not change the supergravity field equations 
$R=0, G_m=0$ by themselves; any arbitrary variation of $R$ or $G_m$, no matter
how complicated it is, will always be multiplied by $R$ or $G_m$ (up to 
derivatives and complex conjugation) and, therefore, will not change any 
solution to the field equations. But that is not the case in the action 
(\ref{rwww}):
a variation of $R$ would be multiplied by $W_{ABC}$ terms which do not vanish 
on-shell and, if it is nontrivial, it could induce changes in the field 
equations. 

Indeed, $R= -3i T_{A \dot A}^{\ \ A \dot A}$ and, in the notation of
our previous paper \cite{m01},
\begin{eqnarray}
\delta T_{A \dot A}^{\ \ \ A \dot A} &=& -\frac{1}{2} H_{A \dot A}^{\ \ \ B 
\dot B}
T_{B \dot B}^{\ \ \ A \dot A} + H^{A B} T_{A \dot A B}^{\ \ \ \ \ \dot A}
+ H^{A \dot B} T_{A \dot A \dot B}^{\ \ \ \ \ \dot A} 
+\frac{1}{2} H^{A B \dot B} T_{A \dot A B \dot B}^{\ \ \ \ \ \ \dot A} 
 \nonumber \\ 
&+& T_{A \dot A}^{\ \ \ A B} H_B^{\ \dot A} +T_{A \dot A}^{\ \ \ A \dot B}
H_{\dot B}^{\ \dot A} - \nabla_{A \dot A} H^{A \dot A} + \nabla^A 
H_{A \dot A}^{\ \ \ \dot A}
\end{eqnarray}
The torsion terms are all proportional to $R$ or $G_m$ and vanish
on-shell. The same is true for the $\nabla_{A \dot A} H^{A \dot A}$ term 
(recall that the full set of $H_M^{\ N}$ was computed in \cite{m01}). The
$\nabla^A H_{A \dot A}^{\ \ \ \dot A}$ term introduces the following 
``dangerous'' terms:
\begin{equation}
\delta R= \nabla^2 \left[ \left(\overline{\nabla}^2 +\frac{1}{3} \overline{R} 
\right) \overline{U} - \frac{3}{4} 
\nabla_{A \dot A} \chi^{A \dot A}- 3i \left( \nabla_A \nabla_{\dot A}-
\nabla_{\dot A} \nabla_A \right) \chi^{A \dot A} \right] + \ldots
\end{equation}
These terms are nonzero on-shell, if (\ref{rwww}) is taken as a correction to 
pure supergravity, and will induce a change in the field equations. This action
has the combined features of all the distinct actions we have analyzed: it
changes the supergravity field equations like (\ref{action}) but, because it 
is proportional to $R$, it can still be reabsorbed in the supergravity action,
by a supervielbein redefinition, like all the actions we have analyzed in this 
section. 
Furthermore, because the supervielbein variations $H_M^{\ N}$ include 
derivatives of $\chi^m$ and $U$, we get, after partial integration, terms with 
derivatives of $R$ and $G_m$ in the superspace field equations, which may mean 
that also with this action the auxiliary fields have nontrivial field 
equations.

After reabsorbing this term, both in pure supergravity and in (\ref{action}), 
the new higher-order superinvariants
which are generated are just powers of the Weyl tensor, which do not vanish on 
shell. Therefore, although this counterterm vanishes on-shell, it generates 
non-trivial higher-order superinvariants, even in pure supergravity 
(without the ${\cal W}^4$ correction in (\ref{action})). 

All the other terms we have considered in this section up to now introduce 
higher powers 
of either ${\cal R}$ or ${\cal R}_{\mu \nu}$, which vanish on-shell if they
are the only quantum corrections to supergravity. They will not vanish on-shell
when we include the ${\cal W}^4$ correction in (\ref{action}). 

We emphasize again that the full loop ($\alpha^\prime$) expansion is {\it not} 
fully given by (\ref{action}), even after having the higher order terms from 
the redefinition of the supervielbein. These terms may need to be included in 
the full action with different numerical coefficients, and also other higher 
order terms, not obtained from the redefinition of the supervielbein, may be 
needed.

\subsection{${\cal W}^4$ terms}
\indent

To conclude our analysis of the ${\cal R}_{\mu \nu \rho \sigma}^4$ 
superinvariants in four dimensions, we discuss now pure Weyl terms.

There are two possible ${\cal W}^4$ polynomials in four dimensions: the
most interesting ${\cal W}_+^2 {\cal W}_-^2$, which we analyzed in \cite{m01}
and in section \ref{2}, and ${\cal W}_+^4+{\cal W}_-^4$.
This last term simply cannot be supersymmetrized, as noticed in \cite{hl81}.
Indeed, in components, this term would be written as 
$\left( \left. \nabla^2 W^2 \right| \right)^2 +{\mathrm h.c.}$, which is not a 
supersymmetric combination (it cannot result from a superspace integration).

%%%%%%%%%%%%%%%%%%%%%%%%%%%%%%%%%%%%%%%%%%%%%%%%%%%%%%%%%%%%%%%%%%%%%%
%%%%%%%%%%%%%%%%%%%%%%%%%%%%%%%%%%%%%%%%%%%%%%%%%%%%%%%%%%%%%%%%%%%%%%

\section{Conclusions}
\label{4}
\indent

From the thirteen independent polynomials that can be built from 
${\cal R}_{\mu \nu \rho \sigma}^4$, only twelve of them can be supersymmetrized
and, from these
twelve, only one - ${\cal W}_+^2 {\cal W}_-^2$ - cannot be reabsorbed in the 
pure supergravity action by a redefinition of the supervielbein. If this term 
is included, the field equations are changed and one gets a nonlocal action 
after elimination of the auxiliary fields.

In this paper, we took the $\alpha {\cal W}_+^2 {\cal W}_-^2$ case which we had
already studied in superspace in a previous paper, and we worked it out in 
$x$-space. We wrote down the action in terms of the components of the $W^2$
superfield. We computed these components, leaving the auxiliary fields which,
on-shell, are an infinite series in the coupling constant $\alpha$. We worked 
these components out completely, just in terms of the vielbein and the 
gravitino, to zeroth order in $\alpha$. We got then the
supersymmetric ${\cal W}_+^2 {\cal W}_-^2$ action, but the field equations
of the auxiliary fields had to be truncated. This should be the three-loop
supergravity effective action, assuming this theory is not finite to this 
order.

We showed that every other ${\cal R}_{\mu \nu \rho \sigma}^4$ term which has a 
${\cal R}$ or ${\cal R}_{\mu \nu}$ factor is supersymmetrizable, but can be 
reabsorbed in the pure supergravity action by a redefinition of 
the supervielbein. For each of these terms, we wrote 
in superspace the respective supersymmetric completion and we proved its 
$R$-invariance. After the supervielbein redefinition, new higher powers of the
Riemann tensor are generated, which we have not analyzed in detail; we argued, 
though, that among these terms should exist powers of the Weyl tensor.

\paragraph{Acknowledgements}
\noindent

The author is grateful to his advisor Martin Ro\v cek and to Ulf Lindstr\"om 
for very helpful discussions. He also wants to thank Peter van Nieuwenhuizen 
for having taught him supergravity and for having allowed him to study through 
the draft of the still unpublished book \cite{west}.

The first part of this work has been supported by Funda\c c\~ao para a 
Ci\^encia e a Tecnologia (Portugal) through grant PRAXIS XXI/BD/11170/97.
The second part has been supported by NSF through grant PHY-0098527.

%%%%%%%%%%%%%%%%%%%%%%%%%%%%%%%%%%%%%%%%%%%%%%%%%%%%%%%%%%%%%%%%%%%%%%
%%%%%%%%%%%%%%%%%%%%%%%%%%%%%%%%%%%%%%%%%%%%%%%%%%%%%%%%%%%%%%%%%%%%%%

\appendix

%%%%%%%%%%%%%%%%%%%%%%%%%%%%%%%%%%%%%%%%%%%%%%%%%%%%%%%%%%%%%%%%%%%%%%
%%%%%%%%%%%%%%%%%%%%%%%%%%%%%%%%%%%%%%%%%%%%%%%%%%%%%%%%%%%%%%%%%%%%%%
\section{From superspace to components in ${\cal N}=1, d=4$ supergravity}
\setcounter{equation}{0}
\label{a1}
\indent

Our conventions have been mostly described in \cite{m01}.
The Riemann tensor admits, in $d$ spacetime dimensions, the following 
decomposition in terms of the Weyl tensor ${\cal W}_{\mu \nu \rho \sigma}$, the
Ricci tensor ${\cal R}_{\mu \nu}$ and the Ricci scalar ${\cal R}$:
\begin{eqnarray}
{\cal R}_{\mu \nu \rho \sigma}&=&{\cal W}_{\mu \nu \rho \sigma}-\frac{1}{d-2}
\left(g_{\mu \rho} {\cal R}_{\nu \sigma} - g_{\nu \rho} {\cal R}_{\mu \sigma} +
g_{\nu \sigma} {\cal R}_{\mu \rho} - g_{\mu \sigma} {\cal R}_{\nu \rho} \right)
\nonumber \\
&+&\frac{1}{(d-1)(d-2)} \left(g_{\mu \rho} g_{\nu \sigma} - g_{\nu \rho} 
g_{\mu \sigma}\right) {\cal R} \label{riemann}
\end{eqnarray}
We define the traceless Ricci tensor as 
\begin{equation}
{\cal S}_{\mu \nu}:= {\cal R}_{\mu \nu} -\frac{1}{d} g_{\mu \nu} {\cal R}
\end{equation}
In four dimensions, the Weyl tensor can still be decomposed in its self-dual 
and antiself-dual parts:
\begin{equation}
{\cal W}_{\mu \nu \rho \sigma}= {\cal W}^+_{\mu \nu \rho \sigma} + 
{\cal W}^-_{\mu \nu \rho \sigma}, {\cal W}^{\mp}_{\mu \nu \rho \sigma}
:=\frac{1}{2} \left({\cal W}_{\mu \nu \rho \sigma} \pm \frac{i}{2}
\varepsilon_{\mu \nu}^{\ \ \ \lambda \tau} {\cal W}_{\lambda \tau \rho \sigma}
\right)
\end{equation}

Superspace supergravity is described by two antichiral superfields
$\overline{R}, W_{ABC}$, their complex conjugates and a real superfield 
$G_{A \dot A}$, which describe the off-shell solution to the Bianchi 
identities. These identities imply the following differential relations 
between the superfields:
\begin{equation}
\nabla^A G_{A\dot B} = \frac{1}{24} \nabla_{\dot B} R  \label{diffg}
\end{equation}

\begin{equation}
\nabla^A W_{ABC} = i \left( \nabla_{B \dot A} G_C^{\ \ \dot A} +\nabla_{C
\dot A} G_B^{\ \ \dot A} \right) 
\label{diffw}
\end{equation}

From (\ref{diffg}) and its complex conjugate and the solution of the Bianchi 
identities, we may also derive the following useful relation between 
superfields:

\begin{equation}
\nabla^2 \overline{R} - \overline{\nabla}^2 R = 96 i \nabla^n G_n 
\label{wb203}
\end{equation}
These relations (\ref{diffg}), (\ref{diffw}), (\ref{wb203}) are off-shell 
identities (not field equations).

In order to determine the component expansion of the supergravity superfields, 
we use the method of gauge completion \cite{west, wz782, wessbagger, ggrs}. 
The basic idea behind it is to relate in superspace some superfields and 
superparameters at $\theta=0$ with some $x$ space quantities, and then to 
require compatibility between the $x$ space and superspace transformation 
rules.

We make the following identification for the supervielbeins at $\theta=0$ 
(symbolically $\left. E_\Pi^{\ N }\right| $:
\begin{equation}
\left. E_\Pi ^{\ \ N}\right| =\left[ \begin{array}{ccc} e_\mu^{\ \ m} &
\frac{1}{2}\psi_\mu^{\ \ A } & \frac{1}{2}\psi_\mu^{\ \ \dot A }\\ 
0 & \delta_B^{\ \ A} & 0 \\ 0 & 0 & \delta_{\dot B}^{\ \ \dot A}
\end{array} \right] \label{vielbeinx}
\end{equation}
In the same way, we gauge the fermionic superconnection at order $\theta=0$
to zero and we can set its bosonic part equal to the usual spin connection: 
\begin{eqnarray}
\left. \Omega_{\mu m}^{\ \ \ n} \right| &=& \omega_{\mu m}^{\ \ \ n}
\left( x\right) \nonumber \\
\left. \Omega_{A m}^{\ \ \ n}\right|, \left. \Omega_{\dot A m}^{\ \ \ n}\right|
 &=&0 \label{connectionx}
\end{eqnarray}

We also identify, at the same order $\theta=0$, the superspace vector 
covariant derivative (with an Einstein indice) with the curved
space covariant derivative:
\begin{equation}
\left. \nabla_\mu \right| = {\cal D}_\mu
\end{equation}
These gauge choices are all preserved by supergravity transformations.

As a careful analysis using the solution to the Bianchi identities and the 
off-shell relations among the supergravity superfields 
$\overline{R}, G_n, W_{ABC}$ shows, the component field content of these 
superfields is all known once we know \footnote{Underlined indices are 
symmetrized with weigth one; undotted and dotted indices are always 
symmetrized independently.}
\begin{equation}
\left. \overline{R} \right|, \left. \nabla_A \overline{R}\right|, \left. \nabla^2 \overline{R}\right|, \left. G_{A \dot A}\right|, \left. 
\nabla_{\underline{A}}G_{\underline{B} \dot A}\right|, \left. 
\nabla_{\underline{\dot A}} \nabla_{\underline{A}} G_{\underline{B} 
\underline{\dot B}}\right|, \left. W_{ABC} \right|, \left. 
\nabla _{\underline{D}}W_{\underline{ABC}}\right| \label{independent}
\end{equation}
and their complex conjugates. 

All the other components and higher derivatives of $\overline{R}, G_{A \dot
A}, W_{ABC}$ can be written as functions of these previous ones. In order to 
determine the ``basic'' components, first we solve for superspace torsions and
curvatures in terms of supervielbeins and superconnections using the 
identifications (\ref{vielbeinx}) and (\ref{connectionx}); then we identify 
them with the off-shell solution to the Bianchi identities. The process is now 
standard \cite{wz782, wessbagger, ggrs, bk}; we simply collect here the 
results.

\begin{eqnarray}
\left. \overline{R}\right| &=& 4 \left( M+iN \right)
\\
\left. G_{A\dot A}\right| &=&\frac{1}{3} A_{A\dot A}
\\
\left. \nabla_A \overline{R} \right| &=& -4\psi_{mn}^{\ \ \ B}
\sigma^{mn}_{\ \ \ AB} -4i \left(M+iN \right) 
\psi_{A \dot B}^{\ \ \ \dot B}-4i A^{m} \psi_{mA}
\\
\left. \nabla_{\underline{\dot C}} G_{A \underline{\dot D}} \right|
&=& \frac{1}{4} \psi_{C \underline{\dot C} \ \ \underline{\dot D} A}^{\ \
\ C} +\frac{i}{6} \left( M+iN \right) \psi_{A \underline{\dot C}
\underline{\dot D}} \nonumber \\ 
&+& \frac{i}{24} \left( 3 A_{A \underline{\dot C}} \psi_{\ \
\underline{\dot D} C}^C -A_{C \underline{\dot C}} \psi_{A \underline{\dot
D}}^{\ \ \ \ \ C} +3 A_{C \underline{\dot C}} \psi_{\ \ \underline{\dot D}
A}^C \right)
\\
\left. W_{ABC} \right| &=&-\frac{1}{4} \psi_{\underline{A} \ \ \underline{B}
\dot C \underline{C}}^{\ \ \dot C} -\frac{i}{4} A_{\underline{A}\ }^{\ \
\dot C} \psi_{\underline{B} \dot C \underline{C}} \nonumber \\
&=& -\frac{1}{4} \psi_{mn \underline{A}} \sigma_{\underline{BC}}^{mn}
-\frac{i}{4} A_m \psi_{n \underline{A}} \sigma_{\underline{BC}}^{mn}
\label{w0}
\end{eqnarray}

With this knowledge, we can already compute $\left. \nabla_m G_n \right|$, 
which is necessary to compute in $x$-space the terms we analyze in the main
text:
\begin{eqnarray}
\left. \nabla_m G_n \right|&=& e_m^{\ \mu} {\cal D}_\mu \left. G_n \right|
-\frac{1}{2} \psi_m^A \left. \nabla_A G_n \right|
-\frac{1}{2} \psi_m^{\dot A} \left. \nabla_{\dot A} G_n \right| \nonumber \\
&=&\frac{1}{3} e_m^{\ \mu} {\cal D}_\mu A_n +\frac{i}{12} \left(M-iN \right) 
\psi_m^A \psi_{n A} -\frac{i}{12} \left(M+iN \right) \psi_m^{\dot A}
\psi_{n \dot A} \nonumber \\
&+&\frac{1}{6} \psi_{np}^{\dot A} \sigma^p_{A \dot A} 
\psi_m^A -\frac{i}{12} A_n \psi_p^{\dot A} \sigma^p_{A \dot A} \psi_m^A 
-\frac{1}{6} \psi_{np}^A \sigma^p_{A \dot A} \psi_m^{\dot A}
-\frac{i}{12} A_n \psi_p^A \sigma^p_{A \dot A} \psi_m^{\dot A} \nonumber \\
&-&\frac{i}{24} \varepsilon_{npqs} \psi^{pq \dot A} \sigma^s_{A \dot A}
\psi_m^A -\frac{1}{24} \varepsilon_{npqs} A^p \psi^{q \dot A} 
\sigma^s_{A \dot A} \psi_m^A \nonumber \\
&-&\frac{i}{24} \varepsilon_{npqs} \psi^{pq A} \sigma^s_{A \dot A} 
\psi_m^{\dot A} +\frac{1}{24} \varepsilon_{npqs} A^p \psi^{q A} 
\sigma^s_{A \dot A} \psi_m^{\dot A}
\end{eqnarray}

To express $\left. \nabla^2 \overline{R}\right|, \left. 
\nabla_{\underline{\dot A}} \nabla_{\underline{A}} G_{\underline{B} 
\underline{\dot B}}\right|, \left. 
\nabla _{\underline{D}}W_{\underline{ABC}}\right|$, one must identify the 
(super)curvature $R_{\mu \nu}^{\ \ \ mn}$ with the $x$-space curvature 
${\cal R}_{\mu \nu}^{\ \ \ mn}$, multiply by the inverse supervielbeins 
$E_M^{\ \mu} E_N^{\ \nu}$, identify with the solution to the Bianchi 
identities for $R_{MN}$ and extract the field contents by convenient index 
contraction/symmetrization. The field content of these components will include
the Riemann tensor in one of its irreducible components (respectively the 
Ricci scalar, the Ricci tensor and the Weyl tensor).

The result for $\left. \nabla^2 \overline{R}\right|$ is well known; it is 
necessary in order to compute the action for pure supergravity 
\cite{wessbagger}:
\begin{eqnarray}
\left. \nabla^2 \overline{R} \right| &=& -8 {\cal R} -\frac{32}{3} 
\left(M^2 + N^2 \right) 
-\frac{16}{3} A^m A_m + 16i e_m^\mu {\cal D}_\mu A^m \nonumber \\
&+& 2 \varepsilon^{mnrs} \sigma_s^{A \dot A} \psi_{mnA} \psi_{r \dot A} - 
2 \varepsilon^{mnrs} \sigma_s^{A \dot A} \psi_{mA} \psi_{nr \dot A} 
\nonumber \\
&+& 8 \left(M+iN \right) \psi_m^{\dot A} \psi^m_{\dot A} + 8 A_n 
\sigma_m^{A \dot A} \psi^m_A \psi^n_{\dot A} - 16i \psi_{mn}^A
\psi^{m \dot A} \sigma^n_{A \dot A}
\end{eqnarray}

The procedure for computing $ \left. 
\nabla_{\underline{\dot A}} \nabla_{\underline{A}} G_{\underline{B} 
\underline{\dot B}}\right|, \left. 
\nabla _{\underline{D}}W_{\underline{ABC}}\right|$ is described in detail 
in appendix B of \cite{t86}, but the full calculation, in terms of the 
supergravity multiplet, is just outlined. The results, expressed as functions 
of the known components, are

\begin{eqnarray}
\left. \nabla_{\underline{A}} \nabla_{\underline{\dot A}} G_{\underline{B} 
\underline{\dot B}}\right| &=& 
-i \left. \nabla_{\underline{A} \underline{\dot A}} 
G_{\underline{B} \underline{\dot B}} \right|
+ \frac{1}{8} {\cal R}_{\mu \nu \rho \sigma}
\sigma^{\mu \nu}_{AB} \sigma^{\rho \sigma}_{\dot A \dot B}
+\left.G_{\underline{A}\underline{\dot A}}\right| \left.
G_{\underline{B}\underline{\dot B}} \right|  
-\frac{1}{48} \sigma^{mn}_{AB} \psi_{m \underline{\dot A}} 
\psi_{n \underline{\dot B}} \left. \overline{R} \right| \nonumber \\
&+& \frac{1}{48} \sigma^{mn}_{\dot A \dot B} \psi_{m \underline{A}} 
\psi_{n \underline{B}} \left. R \right|
+\frac{1}{4} \sigma^{mn}_{AB} \psi_{m \underline{\dot A}} \psi_n^C
\left. G_{C \underline{\dot B}} \right| 
+\frac{1}{4} \sigma^{mn}_{\dot A \dot B} \psi_{m \underline{A}} \psi_n^{\dot C}
\left. G_{\underline{B} \dot C } \right| \nonumber \\
&+&\frac{i}{48} \sigma^m_{\underline{A} \underline{\dot A}} 
\psi_{m \underline{\dot B}} \left.\nabla_{\underline{B}} \overline{R} \right|
+\frac{i}{48} \sigma^m_{\underline{A} \underline{\dot A}} 
\psi_{m \underline{B}} \left.\nabla_{\underline{\dot B}} R \right|
-\frac{i}{8} \sigma^m_{\underline{A} \underline{\dot A}} \psi_m^C 
\left. \nabla_C G_{\underline{B} \underline{\dot B}}\right|  \nonumber \\
&-&\frac{i}{8} \sigma^m_{\underline{A} \underline{\dot A}} \psi_m^C 
\left. \nabla_{\underline{B}} G_{C \underline{\dot B}}\right|
-\frac{i}{8} \sigma^m_{C \underline{\dot A}} \psi_{m \underline{A}}
\left. \nabla^C G_{\underline{B} \underline{\dot B}}\right|
-\frac{i}{8} \sigma^m_{C \underline{\dot A}} \psi_{m \underline{A}}
\left. \nabla_{\underline{B}} G^C_{\ \ \underline{\dot B}}\right| \nonumber \\
&-& \frac{i}{4} \sigma^m_{C \underline{\dot A}} \psi_m^C 
\left. \nabla_{\underline{A}} G_{\underline{B} \underline{\dot B}}\right|
-\frac{i}{8} \sigma^{m \ \dot C}_{\underline{A}} \psi_{m \underline{\dot A}} 
\left. \nabla_{\underline{\dot B}} G_{\underline{B} \dot C}\right|
-\frac{i}{8} \sigma^{m \ \dot C}_{\underline{A}} \psi_{m \underline{\dot A}} 
\left. \nabla_{\dot C} G_{\underline{B} \underline{\dot B}}\right|
\nonumber \\
&+&\frac{i}{4} \sigma^m_{\underline{A} \dot C} \psi_{m}^{\dot C} 
\left. \nabla_{\underline{\dot A}} G_{\underline{B} \underline{\dot B}}\right|
+\frac{i}{8} \sigma^m_{\underline{A} \underline{\dot A}} \psi_{m}^{\dot C}
\left. \nabla_{\underline{\dot B}} G_{\underline{B} \dot C}\right|
+\frac{i}{8} \sigma^m_{\underline{A} \underline{\dot A}} \psi_{m}^{\dot C}
\left. \nabla_{\dot C} G_{\underline{B} \underline{\dot B}}\right|
\nonumber \\
&+&\frac{i}{2} \sigma_{\underline{A}}^{m \ \dot C} \psi_{m \underline{B}}
\left. W_{\dot A \dot B \dot C} \right|
-\frac{i}{2} \sigma^{m C}_{\ \ \ \underline{\dot A}} 
\psi_{m \underline{\dot B}} \left. W_{ABC} \right| \\
\left. \nabla_{\underline{A}} W_{\underline{BCD}} \right| &=&
-\frac{1}{8} {\cal R}_{\mu \nu \rho \sigma}
\sigma^{\mu \nu}_{\underline{AB}} \sigma^{\rho \sigma}_{\underline{CD}} -
\frac{1}{24} \sigma^{mn}_{\underline{AB}} \psi_{m \underline{C}}
\psi_{n \underline{D}} \left. R \right| - \frac{1}{2} 
\sigma^{mn}_{\underline{AB}} \psi_{m}^{\dot C} \psi_{n \underline{C}}
\left. G_{\underline{D} \dot C} \right| \nonumber \\
&-&i \sigma^m_{\underline{A} \dot C} \psi_{m \underline{B}} 
\left. \nabla_{\underline{C}} G_{\underline{D}}^{\ \ \dot C}\right|
+i \sigma^m_{\underline{A} \dot C} \psi_m^{\ \dot C}
\left. W_{\underline{BCD}} \right|
\end{eqnarray}

Due to the identity 
\begin{equation}
\varepsilon_{mnpq} \sigma^{pq}_{AB}=2 i \sigma_{mn AB}
\end{equation}
the operator $\sigma^{\mu \nu}_{\underline{AB}} 
\sigma^{\rho \sigma}_{\underline{CD}}$ acts as an ``antiself-dual projector'':
\begin{eqnarray}
{\cal R}_{\mu \nu \rho \sigma} \sigma^{\mu \nu}_{\underline{AB}} 
\sigma^{\rho \sigma}_{\underline{CD}}&=&{\cal W}^+_{\mu \nu \rho \sigma} 
\sigma^{\mu \nu}_{\underline{AB}} \sigma^{\rho \sigma}_{\underline{CD}}
=:-8{\cal W}_{ABCD} \\
{\cal W}_{\mu \nu \rho \sigma} \sigma^{\mu \nu}_{AB}
\sigma^{\rho \sigma}_{\dot A \dot B} &=& 0 
\end{eqnarray}

One can also show that 
\begin{equation}
{\cal R}_{\mu \nu} \sigma^{\mu}_{A \dot A} \sigma^{\nu}_{B \dot B} =
\sigma^\mu_{\underline{A} \underline{\dot A}} 
\sigma^\nu_{\underline{B} \underline{\dot B}}
\left({\cal R}_{\mu \nu} -\frac{1}{4} g_{\mu \nu} {\cal R} \right)
+\frac{1}{2} \varepsilon_{\dot A \dot B} \varepsilon_{AB} {\cal R}
\end{equation}

Defining ${\cal R}_{A \dot A B \dot B CD}=-\frac{1}{4}
{\cal R}_{A \dot A B \dot B mn} \sigma^{mn}_{CD}$, the decomposition 
(\ref{riemann}) can be written as \cite{bk}
\begin{eqnarray}
{\cal R}_{A \dot A B \dot B CD}&=& 
\frac{1}{8} \varepsilon_{\dot A \dot B} {\cal W}^+_{\mu \nu \rho \sigma}
\sigma^{\mu \nu}_{\underline{AB}} \sigma^{\rho \sigma}_{\underline{CD}}
+\frac{1}{2} \varepsilon_{AB} \sigma^\mu_{\underline{C} \underline{\dot A}} 
\sigma^\nu_{\underline{D} \underline{\dot B}}
\left({\cal R}_{\mu \nu} -\frac{1}{4} g_{\mu \nu} {\cal R} \right) \nonumber \\
&-& \frac{1}{12}
\varepsilon_{\dot A \dot B} \left(\varepsilon_{AC} \varepsilon_{BD} +
\varepsilon_{AD} \varepsilon_{BC} \right) {\cal R} \label{rspinor}
\end{eqnarray}

By extensively (and judiciously) using the equalities 
\begin{eqnarray}
\sigma^m_{C \dot C} \sigma^n_{\underline{A} \underline{\dot A}} 
\sigma^p_{\underline{B} \underline{\dot B}} &=& 
\sigma^n_{C \dot C} \sigma^m_{\underline{A} \underline{\dot A}} 
\sigma^p_{\underline{B} \underline{\dot B}} + \varepsilon_{C \underline{A}}
\sigma^{mn}_{\dot A \dot B} \sigma^p_{\underline{B} \dot C}
- \varepsilon_{\dot C \underline{\dot A}} 
\sigma^{mn}_{AB} \sigma^p_{C \underline{\dot B}} \nonumber \\
&+& 2 \varepsilon_{C \underline{A}} \varepsilon_{\dot C \underline{\dot A}}
\left( \eta^{mp} \eta^{nq} - \eta^{mq} \eta^{np} \right) 
\sigma_{q \underline{B}\underline{\dot B}} \\
\sigma^{mn}_{AB} \sigma^p_{C \dot A} &+& \sigma^{np}_{AB} \sigma^m_{C \dot A} +
\sigma^{pm}_{AB} \sigma^n_{C \dot A} =2i \varepsilon^{mnpq} 
\varepsilon_{\underline{A} C} \sigma_{q \underline{B} \dot A}\\
\sigma^{mn}_{\dot A \dot B} \sigma^p_{A \dot C} &+&
\sigma^{pm}_{\dot A \dot B} \sigma^n_{A \dot C} +
\sigma^{np}_{\dot A \dot B} \sigma^m_{A \dot C} = 2i \varepsilon^{mnpq} 
\varepsilon_{\underline{\dot A} \dot C} \sigma_{q A \underline{\dot B}}
\end{eqnarray}
we carried the full computations of $\left. \nabla_{\underline{A}} 
\nabla_{\underline{\dot A}} G_{\underline{B} \underline{\dot B}}\right|$ and
$\left. \nabla_{\underline{A}} W_{\underline{BCD}} \right|$, the results of 
which we now present (to our knowledge, for the first time in the literature):

\begin{eqnarray}
\left. \nabla_{\underline{A}} \nabla_{\underline{\dot A}} G_{\underline{B} 
\underline{\dot B}}\right| &=& -\frac{1}{2} 
\sigma^\mu_{\underline{A} \underline{\dot A}} 
\sigma^\nu_{\underline{B} \underline{\dot B}} \left(
{\cal R}_{\mu \nu} -\frac{1}{4} g_{\mu \nu} {\cal R} \right) + \frac{1}{9}
A_m A_n \sigma^m_{\underline{A}\underline{\dot A}} 
\sigma^n_{\underline{B}\underline{\dot B}}  - \frac{i}{3} e_m^\mu 
{\cal D}_\mu A_n \sigma^m_{\underline{A}\underline{\dot A}} 
\sigma^n_{\underline{B}\underline{\dot B}}
\nonumber \\
&-& \frac{1}{6} \left(M+iN \right) 
\sigma^m_{\underline{A}\underline{\dot A}} 
\sigma^n_{\underline{B}\underline{\dot B}} \psi_m^{\dot C} \psi_{n \dot C}
-\frac{i}{3} \psi^m_{\underline{A}} \psi_{mn \underline{\dot A}} 
\sigma^n_{\underline{B}\underline{\dot B}}
- \frac{i}{3} \psi_{mn \underline{A}} \psi^m_{\underline{\dot A}} 
\sigma^n_{\underline{B}\underline{\dot B}}
\nonumber \\
&-& \frac{i}{4} \sigma^{mn}_{\dot A \dot B} \sigma^p_{\underline{A} \dot C} 
\psi_{mn \underline{B}} \psi_{p}^{\dot C} + 
\frac{i}{12} \sigma^{np}_{\dot A \dot B} \sigma^m_{\underline{A} \dot C}
\psi_{m \underline{B}} \psi_{np}^{\dot C} +\frac{i}{6} \sigma^{mn}_{AB} 
\sigma^p_{C \underline{\dot A}} \psi_{mn}^C \psi_{p \underline{\dot B}}
\nonumber \\
&+& \frac{1}{12} \varepsilon_{mnpq} \psi^{mn}_{ \underline{A}} 
\psi^p_{\underline{\dot A}} \sigma^q_{\underline{B}\underline{\dot B}}
-\frac{1}{12} \varepsilon_{mnpq} \psi^m_{\underline{A}} 
\psi^{np}_{\underline{\dot A}} \sigma^q_{\underline{B}\underline{\dot B}}
- \frac{i}{48} \varepsilon_{mnpq} A^m \psi^n_{ \underline{A}} 
\psi^p_{\underline{\dot A}} \sigma^q_{\underline{B}\underline{\dot B}}
\nonumber \\
&+& \frac{1}{48}  A_m \sigma^m_{\underline{A}\underline{\dot A}} 
\psi_{n \underline{B}} \psi^n_{\underline{\dot B}} +
\frac{3}{16}  A_m \sigma^n_{\underline{A}\underline{\dot A}} 
\psi_{n \underline{B}} \psi^m_{\underline{\dot B}} -
\frac{7}{48}  A_m \sigma^n_{\underline{A}\underline{\dot A}} 
\psi^m_{\underline{B}} \psi_{n \underline{\dot B}} \nonumber \\
&-& \frac{1}{6} \sigma^{mn}_{AB} \sigma^p_{C \underline{\dot A}} 
A_m \psi_n^C \psi_{p \underline{\dot B}} +\frac{1}{12} 
\sigma^{mp}_{\dot A \dot B} \sigma^n_{\underline{A} \dot C} A_m 
\psi_{n \underline{B}} \psi_{p}^{\dot C} +\frac{1}{4}
\sigma^{mn}_{\dot A \dot B} \sigma^p_{\underline{A} \dot C} A_m 
\psi_{n \underline{B}} \psi_{p}^{\dot C} \nonumber \\
&-& \frac{1}{6} \sigma_m^{C \dot C} \sigma^n_{\underline{A}\underline{\dot A}} 
\sigma^p_{\underline{B}\underline{\dot B}} A^m \psi_{n C} \psi_{p \dot C} 
\label{ddg0} \\
\left. \nabla_{\underline{A}} W_{\underline{BCD}} \right| &=&
-\frac{1}{8} {\cal W}^+_{\mu \nu \rho \sigma}
\sigma^{\mu \nu}_{\underline{AB}} \sigma^{\rho \sigma}_{\underline{CD}} 
-\frac{1}{4} \sigma^{mn}_{\underline{AB}} \sigma^r_{\underline{C} \dot A} A_m 
\psi_{n \underline{D}} \psi_r^{\dot A} 
-\frac{1}{4} \sigma^{mr}_{\underline{AB}} \sigma^n_{\underline{C} \dot A} 
A_m \psi_{n \underline{D}} \psi_r^{\dot A} \nonumber \\
&-&\frac{i}{4} \sigma^{mr}_{\underline{AB}} \sigma^n_{\underline{C} \dot A} 
\psi_{n \underline{D}} \psi_{mr}^{\dot A}
+\frac{i}{4} \sigma^{mn}_{\underline{AB}} \sigma^r_{\underline{C} \dot A} 
\psi_{mn \underline{D}} \psi_r^{\dot A} \label{dw0}
\end{eqnarray}

Knowing these components, we can compute, in $x$-space, any action which 
involves the supergravity multiplet. As an example, necessary for our purposes,
we indicate how to compute some other derivatives of $G_m$ which arise in the 
field content of some of the terms in section \ref{3}.

\begin{eqnarray}
\nabla_{\dot A} G^2 &=& G^{B \dot B} \nabla_{\dot A} G_{B \dot B} \\
\overline{\nabla}^2 G^2 &=& \left( \nabla^{\dot A} G^{B \dot B} \right)
\nabla_{\dot A} G_{B \dot B} -\frac{i}{6} G^{B \dot B} \nabla_{B \dot B}
\overline{R} -\overline{R} G^2 \\
\nabla_A \overline{\nabla}^2 G^2 &=& -2 \left( \nabla^{\dot A} G^{B \dot B} 
\right) \nabla_A \nabla_{\dot A} G_{B \dot B} -\frac{i}{6} 
\left( \nabla_A G^{B \dot B} \right) \nabla_{B \dot B} \overline{R}
-\frac{i}{6} G^{B \dot B} \nabla_A \nabla_{B \dot B} \overline{R} \nonumber \\ 
&-& G^2 \nabla_A \overline{R} - \overline{R} G^{B \dot B} \nabla_A G_{B \dot B}
\\
\nabla^2 \overline{\nabla}^2 G^2 &=& -\frac{1}{1152} \left(\nabla^2 
\overline{R}
\right) \nabla^2 \overline{R} -2 \left(\nabla^A \nabla^{\dot A} G^{B \dot B}
\right) \nabla_{\underline{A}} \nabla_{\underline{\dot A}} G_{\underline{B} 
\underline{\dot B}} -\frac{1}{3} G^2 \nabla^2 \overline{R} \nonumber \\ 
&+& \frac{i}{6} \left( \nabla^{\underline{\dot C}} 
G^{A \underline{\dot D}} \right) \nabla_{A \dot D} \nabla_{\dot C} R
-\frac{i}{3} G^m \nabla_m \nabla^2 \overline{R} - \frac{i}{144} 
\left(\nabla_{\dot C} R \right) \nabla^{A \dot C} \nabla_A \overline{R}
 \nonumber \\
&+& \frac{1}{3456} R \left(\nabla^A \overline{R} \right)
\nabla_A \overline{R} + \frac{1}{96} G^{B \dot B} \left(\nabla_B \overline{R}
\right) \nabla_{\dot B} R + \frac{1}{18} \left(\nabla^m \overline{R} \right) 
\nabla_m R \nonumber \\
&-& \frac{7}{6} G^{B \dot B} \left(\nabla^A \overline{R} \right)
\nabla_{\underline{A}} G_{\underline{B} \dot B} + \frac{7}{12} G_{B \dot B}
\left(\nabla_{\dot A} \overline{R} \right) \nabla^{\underline{\dot A}} 
G^{B \underline{\dot B}} + \frac{2}{9} i R G^m \nabla_m \overline{R} 
\nonumber \\
&-& \frac{i}{3} \overline{R} G^m \nabla_m R - \frac{i}{3} \left(
\nabla^{\underline{A}} G^{\underline{B} \dot B} \right) \nabla_{B \dot B} 
\nabla_A \overline{R} -\frac{5}{3} R \left(\nabla^{\underline{\dot A}} 
G^{B \underline{\dot B}} \right) \nabla_{\dot A} G_{B \dot B} \nonumber \\
&+& \overline{R} R G^2 +8 \left(\nabla^m G^n \right) \nabla_m G_n -8
\left(\nabla^m G^n \right) \nabla_n G_m +8i \varepsilon^{mnrs} 
\left(\nabla_m G_n \right) \nabla_r G_s \nonumber \\
&+& 8i \left(\nabla^{\underline{\dot A}} G^{B \underline{\dot B}} \right)
\nabla^A_{\ \dot A} \nabla_{\underline{A}} G_{\underline{B} \dot B}
-12 \left(\nabla^{\underline{\dot A}} G^{B \underline{\dot B}} \right)
G^A_{\ \dot A} \nabla_{\underline{A}} G_{\underline{B} \dot B} \nonumber \\
&+& 16 W^{\dot A \dot B \dot C} G^B_{\ \dot C} \nabla_{\dot A} G_{B \dot B}
\label{d2g2}
\end{eqnarray}

\section{Detailed calculation of $\left. \nabla_A W^2 \right|$ and 
$\left. \nabla^2 W^2 \right|$ at tree level}
\setcounter{equation}{0}
\label{a2}
\indent
Let's start by computing $\left. \nabla_A W^2 \right|$.
From (\ref{dw2}), we may write
\begin{equation}
\left. \nabla_A W^2 \right|=  - 2 \left. W^{BCD} \right| \left. 
\nabla_{\underline{A}} W_{\underline{BCD}} \right|+{\cal O}\left(\alpha \right)
\end{equation}
which may be written, from (\ref{w0}) and (\ref{dw0}), as

\begin{eqnarray}
\left. \nabla_A W^2 \right| &=& -\frac{1}{16} {\cal R}_{\mu \nu \rho \sigma}
\sigma^{\mu \nu}_{\underline{AB}} \sigma^{\rho \sigma}_{\underline{CD}}
\psi^{rs B} \sigma_{rs}^{CD} -\frac{i}{8} \psi^{rs B} \psi_{mn \underline{D}}
\psi_{p \dot A} \sigma_{rs}^{CD} \sigma^{mn}_{\underline{AB}}
\sigma^{p \ \dot A}_{\ \underline{C}} \nonumber \\
&+& \frac{i}{8} \psi^{rs B} \psi_{p \underline{D}} \psi_{mn \dot A}
\sigma_{rs}^{CD} \sigma^{mn}_{\underline{AB}}
\sigma^{p \ \dot A}_{\ \underline{C}} +{\cal O}\left(\alpha \right)
\label{dw2p}
\end{eqnarray}
Let's consider first the terms in (\ref{dw20}) which depend on the Riemann 
tensor. Using the expansion

\begin{eqnarray}
\sigma^{mn}_{\underline{AB}} \sigma^{pq}_{\underline{CD}} \sigma^{rs AB}&=&
-\frac{2}{3} \sigma^{pq}_{CD} \left(\eta^{ms} \eta^{nr} - \eta^{mr} \eta^{ns}
\right) -\frac{2}{3} \sigma^{mn}_{CD} \left(\eta^{ps} \eta^{qr} - \eta^{pr} 
\eta^{qs} \right) \nonumber \\
&-&\frac{2}{3} \sigma^{rs}_{CD} \left(\eta^{mp} \eta^{nq} 
- \eta^{np} \eta^{mq} \right) -\frac{2}{3}i \varepsilon^{mnrs} \sigma^{pq}_{CD}
-\frac{2}{3}i \varepsilon^{pqrs} \sigma^{mn}_{CD} \nonumber \\
&-&\frac{i}{6} \varepsilon^{pqnr} \sigma^{ms}_{CD} +\frac{i}{6} 
\varepsilon^{pqmr} \sigma^{ns}_{CD} -\frac{i}{6} \varepsilon^{mnqr} 
\sigma^{ps}_{CD} +\frac{i}{6} \varepsilon^{mnpr} \sigma^{qs}_{CD} \nonumber \\
&+&\frac{i}{6} \varepsilon^{pqns} \sigma^{mr}_{CD} -\frac{i}{6} 
\varepsilon^{pqms} \sigma^{nr}_{CD} +\frac{i}{6} \varepsilon^{mnqs} 
\sigma^{pr}_{CD} -\frac{i}{6} \varepsilon^{mnps} \sigma^{qr}_{CD} \nonumber \\
&+&\frac{i}{6} \sigma^{\ s}_{u \ CD} \left( \varepsilon^{pqmu} \eta^{nr}
- \varepsilon^{pqnu} \eta^{mr} + \varepsilon^{mnpu} \eta^{qr}
- \varepsilon^{mnqu} \eta^{pr} \right) \nonumber \\
&-&\frac{i}{6} \sigma^{\ r}_{u \ CD} \left( \varepsilon^{pqmu} \eta^{ns}
- \varepsilon^{pqnu} \eta^{ms} + \varepsilon^{mnpu} \eta^{qs}
- \varepsilon^{mnqu} \eta^{ps} \right) \label{sss}
\end{eqnarray}
those terms can be obtained from the corresponding one in (\ref{dw2p}).
The terms in (\ref{dw20}) which do not depend on the Riemann tensor can also 
be obtained from (\ref{dw2p}) using the expansion

\begin{eqnarray}
\sigma^{mn}_{\underline{AB}} \sigma^{rs CD} \chi_{\underline{C}}
\psi_{\underline{D}}&=&\frac{1}{6} \sigma^{mn}_{AB} \sigma^{rs CD} \chi_C
\psi_D - \left(\eta^{ms} \eta^{nr} - \eta^{mr} \eta^{ns}
+i \varepsilon^{mnrs} \right) \chi_{\underline{A}} 
\psi_{\underline{B}}\nonumber \\
&+&\frac{1}{6} \sigma^{ms \ D}_{\ \ \underline{A}} \eta^{nr} 
\chi_{\underline{B}} \psi_D - \frac{1}{6} \sigma^{mr \ D}_{\ \ \underline{A}} 
\eta^{ns} \chi_{\underline{B}} \psi_D + \frac{1}{6} 
\sigma^{ms \ D}_{\ \ \underline{A}} \eta^{nr} \chi_D \psi_{\underline{B}}
\nonumber \\
&-& \frac{1}{6} \sigma^{mr \ D}_{\ \ \underline{A}} \eta^{ns} \chi_D 
\psi_{\underline{B}}
- \frac{1}{6} \sigma^{ns \ D}_{\ \ \underline{A}} \eta^{mr}
\chi_{\underline{B}} \psi_D
+ \frac{1}{6} \sigma^{nr \ D}_{\ \ \underline{A}} \eta^{ms}
\chi_{\underline{B}} \psi_D
\nonumber \\
&-&\frac{1}{6} \sigma^{ns \ D}_{\ \ \underline{A}} \eta^{mr} \chi_D 
\psi_{\underline{B}} + \frac{1}{6} \sigma^{nr \ D}_{\ \ \underline{A}} 
\eta^{ms} \chi_D \psi_{\underline{B}} +\frac{i}{6} 
\varepsilon^{mnr}_{\ \ \ \ t}
\sigma^{ts \ D}_{\ \ \underline{A}} \chi_{\underline{B}} \psi_D \nonumber \\
&-&\frac{i}{6} \varepsilon^{mns}_{\ \ \ \ t}
\sigma^{tr \ D}_{\ \ \underline{A}} \chi_{\underline{B}} \psi_D
+ \frac{i}{6} \varepsilon^{mnr}_{\ \ \ \ t}
\sigma^{ts \ D}_{\ \ \underline{A}} \chi_D \psi_{\underline{B}}
- \frac{i}{6} \varepsilon^{mns}_{\ \ \ \ t}
\sigma^{tr \ D}_{\ \ \underline{A}} \chi_D \psi_{\underline{B}} \nonumber \\
\end{eqnarray}
This concludes our derivation of (\ref{dw20}).

In order to compute $\left. \nabla^2 W^2 \right|$ we write, from (\ref{d2w2}),
\begin{equation}
\left. \nabla^2 W^2 \right|= -2 \left.\nabla^A W^{BCD} \right| 
\left. \nabla_{\underline{A}} W_{\underline{BCD}} \right|
+{\cal O}\left(\alpha \right)
\end{equation}
which we may write, from (\ref{dw0}), as
\begin{eqnarray}
\left. \nabla^2 W^2 \right|&=&-2 {\cal W}^{ABCD} {\cal W}_{ABCD} +
\frac{i}{8} {\cal R}_{\mu \nu \rho \sigma} \psi^{mD} \psi_{rs \dot A}
\sigma^{C \dot A}_m \sigma^{rs AB} \sigma^{\mu \nu}_{\underline{AB}}
\sigma^{\rho \sigma}_{\underline{CD}} \nonumber \\
&-&\frac{i}{8} {\cal R}_{\mu \nu \rho \sigma} \psi_{rs}^D \psi^m_{\dot A}
\sigma^{C \dot A}_m \sigma^{rs AB} \sigma^{\mu \nu}_{\underline{AB}}
\sigma^{\rho \sigma}_{\underline{CD}} +\frac{1}{8}\sigma^{mn}_{\underline{AB}}
\sigma^{pqAB} \sigma^{r\ \dot A}_{\underline{C}} \sigma^{s C \dot B}
\psi_{r \underline{D}} \psi_{mn \dot A} \psi_s^D \psi_{pq \dot B} \nonumber \\
&+& \frac{1}{8}\sigma^{mn}_{\underline{AB}} \sigma^{pqAB} 
\sigma^{r\ \dot A}_{\underline{C}} \sigma^{s C \dot B} \psi_{mn \underline{D}} 
\psi_{r \dot A} \psi_{pq}^D \psi_{s \dot B} \nonumber \\
&-& \frac{1}{4} \sigma^{mn}_{\underline{AB}} \sigma^{pqAB} 
\sigma^{r\ \dot A}_{\underline{C}} \sigma^{s C \dot B} \psi_{r \underline{D}} 
\psi_{mn \dot A} \psi_{pq}^D \psi_{s \dot B} +{\cal O}\left(\alpha \right)
\end{eqnarray}
From the decomposition (\ref{rspinor}) one can easily derive 
\begin{equation}
{\cal W}^{ABCD} {\cal W}_{ABCD}= {\cal W}_+^{\mu \nu \rho \sigma}
{\cal W}^+_{\mu \nu \rho \sigma}
\end{equation}
Using the expansion (\ref{sss}), it is straightforward to check that 
\begin{eqnarray}
{\cal R}_{\mu \nu \rho \sigma} \psi^{mD} \psi_{rs \dot A}
\sigma^{C \dot A}_m \sigma^{rs AB} \sigma^{\mu \nu}_{\underline{AB}}
\sigma^{\rho \sigma}_{\underline{CD}}&=& \frac{64}{3} 
{\cal W}^+_{\mu \nu \rho \sigma} \psi^{\rho A} \psi^{\mu \nu \dot A}
\sigma_{A \dot A}^{\sigma} \nonumber \\
{\cal R}_{\mu \nu \rho \sigma} \psi_{rs}^D \psi^m_{\dot A}
\sigma^{C \dot A}_m \sigma^{rs AB} \sigma^{\mu \nu}_{\underline{AB}}
\sigma^{\rho \sigma}_{\underline{CD}}&=& \frac{64}{3} 
{\cal W}^+_{\mu \nu \rho \sigma} \psi^{\mu \nu A} \psi^{\rho \dot A}
\sigma_{A \dot A}^{\sigma}
\end{eqnarray}
Having these expressions, we get (\ref{d2w20}). 

We include here the full
$\sigma$-matrix expansion for the four-fermion terms; we did not include it in 
the main body to keep the text more readable:
\begin{eqnarray}
&\sigma^{mn}_{AB}& \sigma^{r\ \dot A}_C \sigma^{pq \underline{AB}}
\sigma^{s \underline{C} \dot B} \varepsilon^{\underline{D}E} \nonumber \\
&=& \frac{1}{12} \sigma^{mn}_{AB} \sigma^{pqAB} \sigma^{r\ \dot A}_C 
\sigma^{s C \dot B} \varepsilon^{DE} +\frac{1}{12} \sigma^{mn}_{AB} 
\sigma^{pqAB} \sigma^{r\ \dot A}_C \sigma^{s D \dot B} \varepsilon^{CE}
\nonumber \\
&+& \frac{1}{6} \sigma^{mn}_{AB} \sigma^{pqAC} \sigma^{r\ \dot A}_C
\sigma^{s B \dot B} \varepsilon^{DE} +\frac{1}{6} \sigma^{mn}_{AB} 
\sigma^{pqAD} \sigma^{r\ \dot A}_C \sigma^{s C \dot B} \varepsilon^{BE} 
\nonumber \\
&+& \frac{1}{6} \sigma^{mn}_{AB} \sigma^{pqCD} \sigma^{r\ \dot A}_C
\sigma^{s B \dot B} \varepsilon^{AE} + \frac{1}{6} \sigma^{mn}_{AB} 
\sigma^{pqAC} \sigma^{r\ \dot A}_C \sigma^{s D \dot B} \varepsilon^{BE}
\nonumber \\
&+& \frac{1}{6} \sigma^{mn}_{AB} \sigma^{pqAD} \sigma^{r\ \dot A}_C
\sigma^{s B \dot B} \varepsilon^{CE} \nonumber \\
&=&-\frac{1}{3} \left(\eta^{mq} \eta^{np} - \eta^{mp} \eta^{nq} 
+i \varepsilon^{mnpq} \right) \left(\sigma^{rs \dot A \dot B} +\eta^{rs}
\varepsilon^{\dot A \dot B} \right) \varepsilon^{DE} \nonumber \\
&+& \frac{1}{6} \left(\eta^{mq} \eta^{np} - \eta^{mp} \eta^{nq} 
+i \varepsilon^{mnpq} \right) \sigma^{r E \dot A}
\sigma^{s D \dot B} \nonumber \\
&-& \frac{1}{6} \left(\eta^{np} \sigma^{mq DE} - \eta^{mp} \sigma^{nq DE} 
+i \varepsilon^{mnpt} \sigma_t^{\ q DE} \right)  \left(\eta^{rs}
\varepsilon^{\dot A \dot B} +\sigma^{rs \dot A \dot B} \right) \nonumber \\
&-& \frac{1}{6} \left(\eta^{mp} \eta^{ns} \eta^{rq} - \eta^{np} \eta^{ms}
\eta^{rq} - \eta^{mq} \eta^{ns} \eta^{rp} + \eta^{nq} \eta^{ms}
\eta^{rp}  \right) \varepsilon^{\dot A \dot B} \varepsilon^{DE} \nonumber \\
&-&  \frac{i}{6} \left(\varepsilon^{mnsp} \eta^{rq} - \varepsilon^{mnsq} 
\eta^{rp} +\varepsilon^{pqrm} \eta^{ns} -\varepsilon^{pqrn} \eta^{ms}
 \right) \varepsilon^{\dot A \dot B} \varepsilon^{DE} \nonumber \\
&-& \frac{1}{6} \left(\eta^{ms} \eta^{rq} \sigma^{np \dot A \dot B} 
+ \eta^{ns} \eta^{rp} \sigma^{mq \dot A \dot B} 
- \eta^{ns} \eta^{rq} \sigma^{mp \dot A \dot B} 
- \eta^{ms} \eta^{rp} \sigma^{nq \dot A \dot B} \right) \varepsilon^{DE} 
\nonumber \\
&-& \frac{i}{6} \left(\varepsilon^{mnsu} \eta^{rp} \sigma_u^{\ q \dot A \dot B}
- \varepsilon^{mnsu} \eta^{rq} \sigma_u^{\ p \dot A \dot B}
+ \varepsilon^{pqru} \eta^{ms} \sigma_u^{\ n \dot A \dot B} 
- \varepsilon^{pqru} \eta^{ns} \sigma_u^{\ m \dot A \dot B} \right) 
\varepsilon^{DE} \nonumber \\
&+& \frac{1}{6} \left(\eta^{mu} \eta^{ns} - \eta^{ms} \eta^{nu} +i
\varepsilon^{mnsu} \right) \left(\eta^{pv} \eta^{rq}- \eta^{rp} \eta^{qv}
+ i \varepsilon^{pqrv} \right) \sigma_u^{E \dot A}
\sigma_v^{D \dot B} \nonumber \\
&+& \frac{1}{6} \left[ \left(\eta^{mq} \eta^{rp} - \eta^{mp} \eta^{rq} +i
\varepsilon^{mpqr} \right) \sigma^{n E \dot A} -
\left(\eta^{nq} \eta^{rp} - \eta^{np} \eta^{rq} +i
\varepsilon^{npqr} \right) \sigma^{m E \dot A} \right]
\sigma^{s D \dot B} \nonumber \\
&+& \frac{1}{6} \left[ \left(\eta^{ms} \eta^{np} - \eta^{mp} \eta^{ns} +i
\varepsilon^{mnps} \right) \sigma^{q D \dot B} -
\left(\eta^{nq} \eta^{ms} - \eta^{ns} \eta^{mq} +i
\varepsilon^{mnqs} \right) \sigma^{p D \dot B} \right]
\sigma^{r E \dot A} \nonumber \\
&+& \frac{i}{6} \left(\varepsilon^{mnpu} \eta^{qr} - \varepsilon^{mnqu} 
\eta^{pr} \right) \sigma_u^{E \dot A} \sigma^{s D \dot B}
+ \frac{i}{6} \left(\varepsilon^{mpqu} \eta^{ns} - \varepsilon^{npqu} 
\eta^{ms} \right) \sigma^{r E \dot A} \sigma_u^{D \dot B} \nonumber \\
&+& \frac{1}{6} \varepsilon^{mnuv} \varepsilon^{pqr}_{\ \ \ v}
\sigma_u^{E \dot A} \sigma^{s D \dot B} + \frac{1}{6} \varepsilon^{mnsv} 
\varepsilon^{pqu}_{\ \ \ v} \sigma^{r E \dot A} \sigma_u^{D \dot B}
\nonumber \\
&+& \frac{1}{6} \varepsilon^{mnsu} \varepsilon^{pqrv} \left(\eta_{uv} 
\varepsilon^{\dot A \dot B} - \sigma_{uv}^{\dot A \dot B} \right) 
\varepsilon^{DE} 
\end{eqnarray} 

%%%%%%%%%%%%%%%%%%%%%%%%%%%%%%%%%%%%%%%%%%%%%%%%%%%%%%%%%%%%%%%%%%%%%%
%%%%%%%%%%%%%%%%%%%%%%%%%%%%%%%%%%%%%%%%%%%%%%%%%%%%%%%%%%%%%%%%%%%%%%

\end{document}